\documentclass[useAMS,usenatbib,usegraphicx]{mn2e}
\usepackage{mathtext,amssymb}
\usepackage{epsfig}
\usepackage{url}
\usepackage{natbib}
\usepackage{color}
\usepackage{rotating}
\usepackage{pdflscape}
\usepackage[english]{babel} 
\usepackage[koi8-r]{inputenc}       
\usepackage{hyperref}
\newcommand{\kms}{\ensuremath{\mbox{km~s}^{-1}}}
\newcommand{\kpc}{\ensuremath{\rm kpc}}
\newcommand{\Msun}{\ensuremath{\rm M_\odot}}
\newcommand{\Lsun}{\ensuremath{\rm L_\odot}}

   \newcommand{\aj}{AJ}                      
   
   \newcommand{\araa}{ARA\&A}                  
   \newcommand{\apj}{ApJ}                      
   \newcommand{\apjs}{ApJS}                    
   \newcommand{\ao}{Applied Optics}
   
   \newcommand{\aap}{A\&A}                      
   
   \newcommand{\aaps}{A\&AS}                      

   \newcommand{\mnras}{MNRAS}

   \newcommand{\pasp}{Publications of the ASP}

\title[High reddening of MT304]{On the nature of high reddening of Cygnus~OB2~\#12 hypergiant}

\author[Maryeva et al.]{O.~V.~Maryeva$^{1}$\thanks{E-mail:
olga.maryeva@gmail.com},  E.~L.~Chentsov$^1$,  V.~P.~Goranskij$^2$,  V.~V.~Dyachenko$^1$, S.~V.~Karpov$^{1,3}$, \\
                 \\
{\LARGE \rm E.~V. Malogolovets$^1$, D.~A.~Rastegaev$^1$ }\\
\\
$^{1}$Special Astrophysical Observatory of the Russian Academy of Sciences, Nizhnii Arkhyz, 369167, Russia\\
$^{2}$Sternberg Astronomical Institute, Moscow State University, Universitetsky pr., 13, Moscow, 119992, Russia\\
$^{3}$Astronomy and  geodesy department of Kazan (Volga region) Federal University,Kremlevskaya str., 18, Kazan, 420008, Russia\\
}

\begin{document}
\date{Accepted {}, Received {}, in original form {}}

\pagerange{\pageref{firstpage}--\pageref{lastpage}} \pubyear{2014}

\maketitle

\label{firstpage}

\begin{abstract}

         To explain the nature of the high reddening ($A_V\simeq 10$~mag) towards one of the most luminous stars in the Galaxy -- Cyg~OB2~\#12 (B5 Ia-0), also known as MT304,  we carried out spectro-photometric observations of 24 stars located in its vicinity.   We included  five new B-stars {\color{black} among} the members of Cygnus OB2, and for five more photometrically selected  {\color{black} stars } we spectroscopically confirmed their membership.
         We constructed the map of interstellar extinction within 2.5~arcmin radius  and found that interstellar extinction increases towards  MT304. According to our results the most reddened  OB-stars in the association after  MT304 are J203240.35+411420.1 and J203239.90+411436.2, located about 15 arcsec away from it. Interstellar extinction towards these stars is about 9~mag.  The increase of reddening towards MT304 suggests that the reddening excess may be caused by the circumstellar shell ejected by the star during its evolution. This shell  absorbs 1~mag, but its chemical composition and temperature are unclear.  We also report the  detection of  a second component of MT304, and discovery of  an even fainter third component, based  on data of speckle  interferometric observations taken with the Russian 6-m telescope.


\end{abstract}


\section{Introduction}\label{sec:intro}

            Stellar association Cyg~OB2 (VI~Cyg) is one of the closest star formation regions to the Sun (distance is  $1.4\pm0.1$~kpc \citep{Rygl}) which  was discovered by \citet{MunchMorgan} in the middle of the last century. As of now Cyg~OB2 is one of the richest OB associations known in our  Galaxy. According to \citet{Wright2015} it consists of at least 169 OB-stars, and its total mass is 16~500$\,\Msun$. A zoo of unique objects is assembled in Cyg~OB2, such as two stars of   early spectral class O3f (Cyg~OB2~\#7\footnote{Here and below the notation Cyg~OB2~\#N means the name in the catalog by \citet{Schulte58}.} and \#22), four stars  are Ofc objects (Cyg~OB2~\#9, \#11, \#8A  and \#8C), and there is also an {\it enigmatic} \ star Cyg~OB2~\#12 (Schulte~12, MT304). MT304\footnote{MT is the catalogue of Cyg~OB2 members by \citet{MT91}} is  {\it enigmatic} not only due to its high luminosity (its bolometric luminosity is $1.9\cdot 10^6 \Lsun$ \citep{Clark} and according to various estimates, the star is one of the brightest stars in the Galaxy \citep{Clark}), but also because of its strong reddening, which is higher than average reddening of the massive stars in the association (see \citet{Chentsov} and references therein, and \citet{Wright2015}).

            MT304 is classified as  B5 Ia-0 \citep{Chentsov} or B3-4 Ia \citep{Clark} hypergiant. Since the star lies on the Hertzsprung-Russell (H-R) diagram higher than the Humphreys-Davidson (H-D) limit, it is included in the list of Galactic candidate Luminous Blue Variables (cLBV) \citep{Clark2005,VinkEtaCar}; however, it lacks the characteristic brightness variability an LBV should possess. Until now only the changes of spectral line profiles  with time are detected \citep{Chentsov}. \citet{Clark} analyzed spectral and photometric observations of MT304, obtained since 1954 and concluded that flux and spectral appearance of MT304 are relatively constant, and they classified the star as a blue hypergiant (BHG).

\begin{table}
\caption{Observations log. $\Delta\lambda $ is the spectral resolution in 4000--5700~\AA. In lower table $N_{MT}$ is the number of non-saturated and measured accurately enough stars from \citet{MT91} we used for photometric calibration. The last column shows the mean photometric error of stars measured on given image.} 
\label{tab:log}
\vspace{0.5cm}
\begin{tabular}{l|c|c|c|c}
\hline 
\hline
\multicolumn{5}{c}{Spectroscopy with SCORPIO on 6-m telescope}                            \\ 
                   &  Seeing         & \multicolumn{1}{c}{  $\Delta\lambda $}   & \multicolumn{2}{c}{  Spectral}   \\ 
     Date          &[$\arcsec{\,}$~] & \multicolumn{1}{c}{  [\AA] }             & \multicolumn{2}{c}{  standard }  \\ 
                   &                 &                                          & \multicolumn{2}{c}{  star     }  \\ 
 2013 May 30       &  1.1-1.3        &6.4 & \multicolumn{2}{c}{BD33d2642  }  \\ 
 2014 Mar 7        &     3.2         &6.6 & \multicolumn{2}{c}{  HZ44     }  \\ 
 2014 Mar 8        &     3.0         &6.0 & \multicolumn{2}{c}{BD75d325   }  \\ 
 2014 Jul 31       &  1.2-1.5        &7.4 & \multicolumn{2}{c}{BD28d4211  }  \\ 
 2014 Aug 3        &  1.2-1.5        & 8  & \multicolumn{2}{c}{BD33d2642  }  \\ 
\hline 
\hline                  
 \multicolumn{5}{c}{Photometry with SCORPIO on 6-m telescope}                         \\ 
                   &   Filter        &    Exp.                 &  $N_{MT}$  &         \\ 
                   &                 &Time [s]                 &            &     \\ 
 2013 May 30       &    B            &      3                  & 18         &  0.02   \\
                   &    B            &      30                 & 18         &  0.05   \\
                   &    V            &      1                  & 18         &  0.01   \\ 
                   &    V            &      30                 & 12         &  0.01   \\
 \multicolumn{5}{c}{ }                                                      \\ 
  \multicolumn{5}{c}{Photometry on Zeiss-1000}                              \\ 
 2012 Oct 14       &    B            &    200                  &  33        &  0.03   \\
                   &    V            &    120                  &  31        &  0.02   \\ 
\hline                   
\end{tabular}
\end{table}
 
            Interstellar extinction towards Cyg~OB2 was investigated in some works \citep{Hanson2003,Wright}. \citet{KiminkiAv} determined the extinction for all bright ($V<15$~mag) stars belonging to association and found that extinction is inhomogeneous with $A_V$ varying between 3.5 and  7.7~mag  (except for MT~304). \citet{Drew} using data from the INT/WFC Photometric H$\alpha$ Survey (IPHAS) found clustered  A-type stars at distance of 20~arcmin south of the centre of Cyg~OB2 and estimated extinction for these stars $4.5<A_V<7$~mag. IPHAS data also were used by  \citet{Sale2009} for constructing  a 3D map of Galactic extinction. They  show that towards  Cyg~OB2 at distance of 1~kpc $A_V$ reaches about 2~mag and then rapidly increases up to 6~mag at 2~kpc. \citet{Guarcello} supposing that the age of the association is 3.5~Myr constructed the map of reddening with angular resolution of 3~arcmin based on multicolor photometry. Using $r-i$ versus $i-z$ color-color diagram they found that extinction is between 2~mag and 6~mag, and mean extinction in the center of Cyg~OB2 is $A_V=4.33$~mag. Recently \citet{Wright2015} estimated mean extinction for stars belonging to the association as $A_V=5.4$~mag, while for a half  of stars  $A_V$ is in range 4.5-6.7~mag.

            Compared to other bright massive stars in the association  MT304 is significantly more reddened. At first \citet{Sharpless} estimated interstellar extinction towards it as  $A_V\simeq10.1$~mag. This estimate is in good agreement with modern data \citep{KiminkiAv,Wright2015}. According to \citet{Wright2015} the difference between MT304 and MT488, second most reddened OB-star, is $\Delta A_V=1.9$~mag. Nature of this excessive absorption remains unclear. Does it originate in a small dense dust cloud which is accidentally
            caught in the line of sight (see \citet{Whittet2015} and discussion therein)? Or does it arise in a circumstellar shell (see for example \citet{Chentsov})?
            \citet{Clark}, who analysed MT304 spectrum using {\sc cmfgen} atmospheric code   \citep{Hillier5}, did not find any evidence of an infrared emission of dust with temperatures typical for LBV ejected shells, and therefore the circumstellar origin of the reddening is questionable. 
           

           Recently \citet{Caballero} reported the discovery of a second component of MT304 based on observations of the Hubble Space Telescope (HST). The distance between the components is $63.6~{\rm mas}$ and the magnitude difference is $\Delta$m=$2.3\pm0.2$~mag {\color{black} in F583W filter}. Such a fainter component  can not significantly bias the reddening estimation of MT304. 
                                 
           In order to investigate the extinction near MT304 we carried out long-slit spectroscopy and photometry of stars  within  2.5~arcmin  from it. Moreover we refined parameters of the second companion of MT304 using speckle-interferometer observations with the Russian 6-meter telescope. In this article we will report the results of these observations. 
           
           This paper is organized as follows. The observational data  and data reduction process are described in the next section. Results are presented in Section~\ref{sec:res}, and are discussed in Section~\ref{sec:disc}.  Finally, in Section~\ref{sec:concl} we summarize the main points of our work. 
           
\begin{figure}
\includegraphics[scale=0.7,viewport=0 0 340 340,clip]{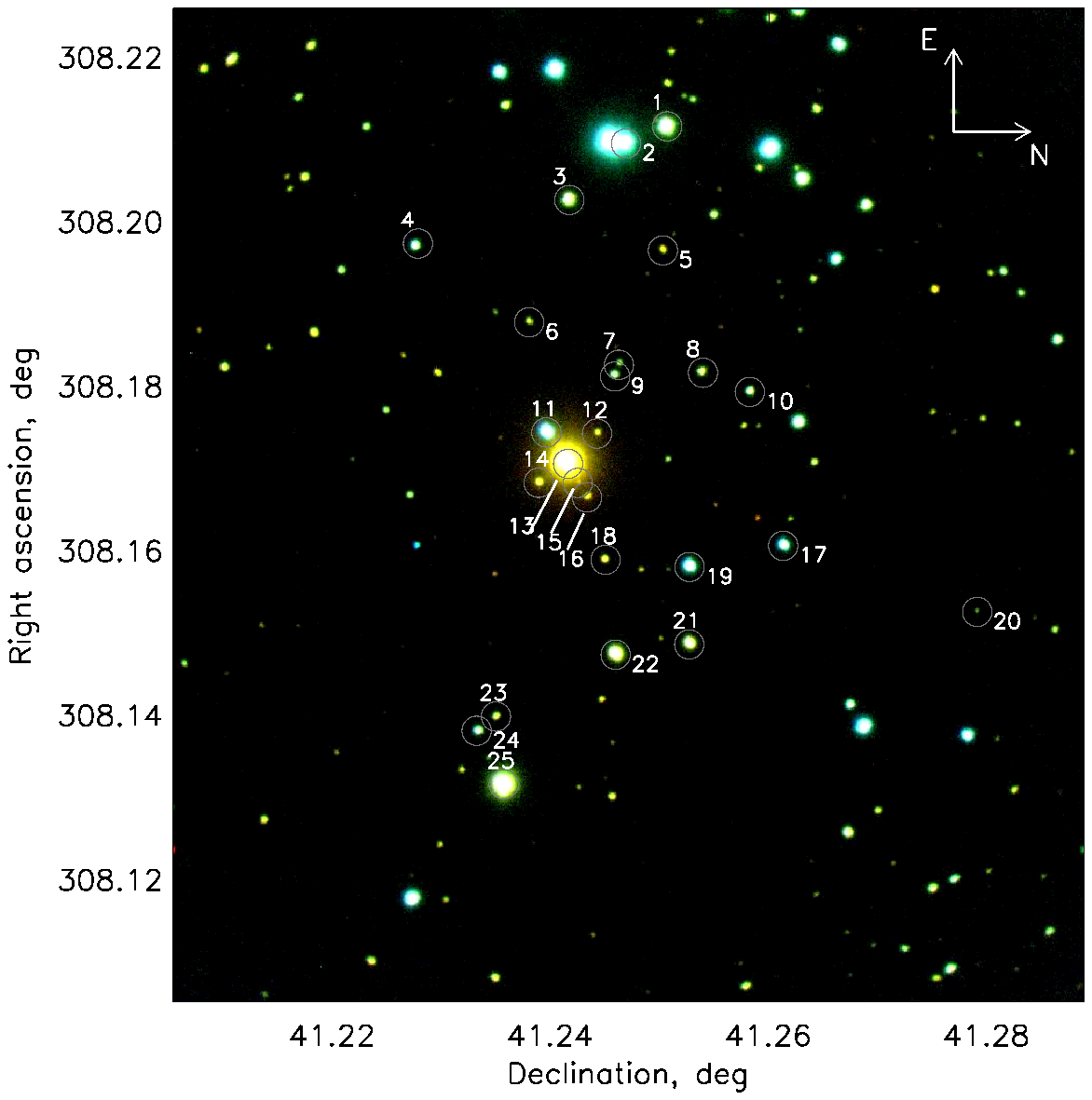}
\caption{Color image (B, V and R bands) of the  central part of Cyg~OB2 close to MT304, obtained with SCORPIO. Circles mark the stars studied in this work.}
\label{fig:rgbfirst}
\includegraphics[scale=0.7,viewport=0 0 322 340,clip]{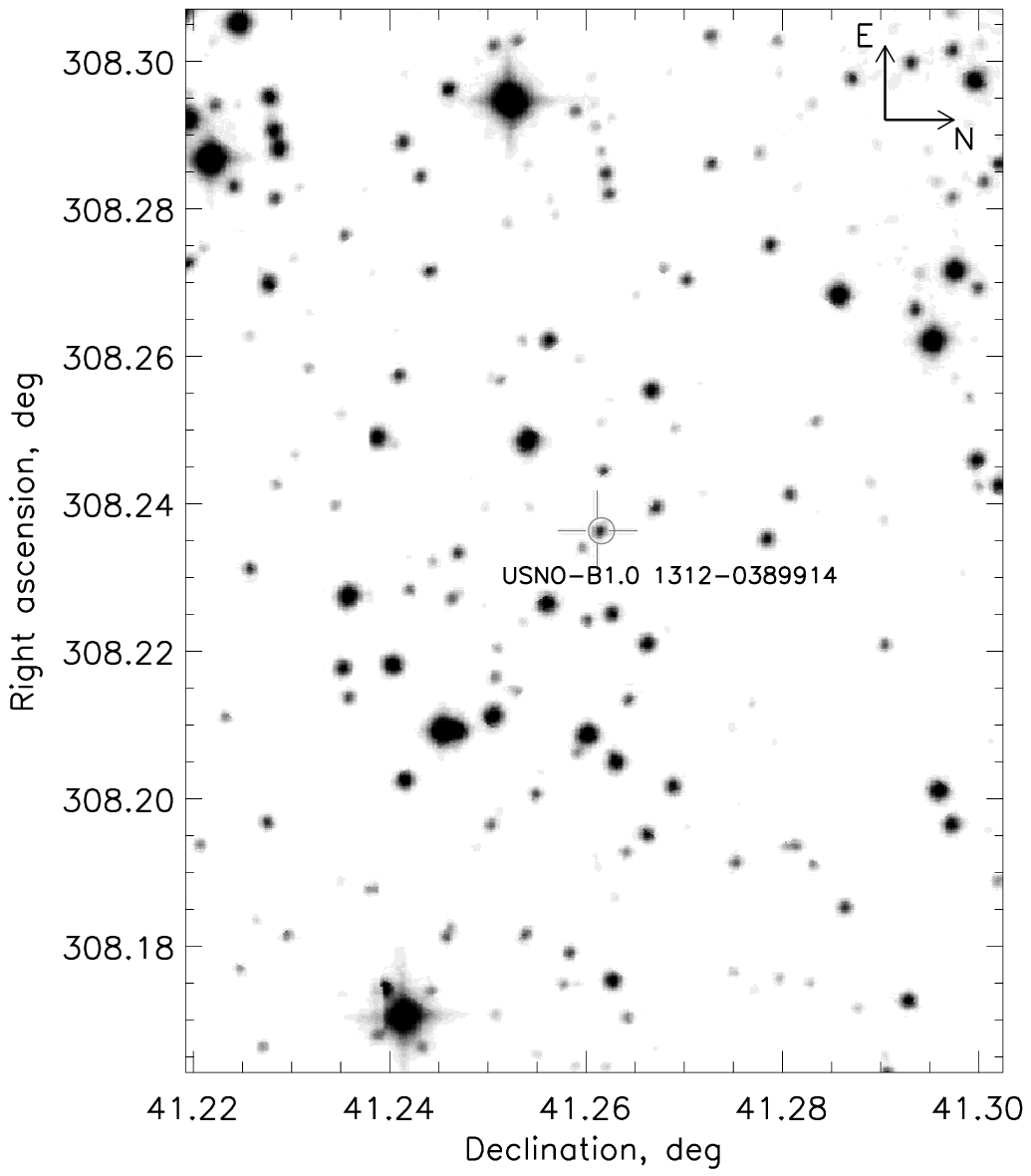} 
\caption{Identification chart of the star USNO-B1.0 1312-0389914 from DSS.}
\label{fig:rgb}
\end{figure}
\section{Observations}\label{sec:obs}

\subsection{Long-slit spectroscopy}\label{sec:longslit}

           Longslit spectroscopy of stars close to MT304 was carried out with the Russian 6-m telescope with the Spectral Camera with Optical  Reducer for Photometric and Interferometric Observations (SCORPIO) \citep{scorpio} during five nights in 2013-2014, using the grism VPHG~1200G. Also, every night the spectrum of a spectrophotometric standard star from the list of \citet{oke} has been obtained  for flux-calibration. Names of standard stars are given in the last column of Table~\ref{tab:log}. Table~\ref{tab:log} lists the dates of the observations and seeings, as well as the spectral resolution  estimated for every night using the spectrum of He-Ar-Ne lamp.
            
         Figure~\ref{fig:rgbfirst} shows area around MT304, the stars observed spectroscopically are marked by circles and are numbered except for USNO-B1.0 1312-0389914 ($\alpha=20:32:56.696, \delta=+41:15:41.26$); the identification chart for the latter object is given separately in the Figure~\ref{fig:rgb}. 
           Table~\ref{tab:names}  provides  the catalog names of these stars and their numbers in Figure~\ref{fig:rgbfirst}.

           All the SCORPIO spectra were reduced using the {\tt ScoRe} package\footnote{{\tt ScoRe} package \url{http://www.sao.ru/hq/ssl/maryeva/score.html}}. {\tt ScoRe} was written by Maryeva and Abolmasov in IDL language for SCORPIO long-slit data reduction.  Package incorporates all the standard stages of long-slit data reduction process. The resulting  spectra in flux units are shown in Figures~\ref{fig:spectrum17}-\ref{fig:spectrumusno} in the Appendix.

\subsection{Photometry}\label{sec:photometry}

\begin{table}
\caption{The sample of stars studied in this work. N is  labels according to Figure~\ref{fig:rgb}, SDSS -- name in Sloan Digital Sky Survey (SDSS) 
         catalog,  MT/AFM  -- name in catalog by \citet{MT91} or \citet{AFM}.} 
\label{tab:names}
\vspace{0.5cm}
\begin{tabular}{l|l|l}
\hline
N   &      SDSS               &  MT/AFM        \\
\hline
 1  & J203250.75+411502.2     &  MT343         \\    
 2  & J203250.25+411448.4     &  MT340         \\    
 3  & J203248.62+411429.8     &  MT333         \\    
 4  & J203247.28+411339.0     &                \\    
 5  & J203247.17+411501.0     &                \\    
 6  & J203245.07+411416.5     &  AFM91         \\    
 7  & J203243.84+411446.5     &                \\    
 8  & J203243.61+411513.9     &  AFM79         \\    
 9  & J203243.49+411445.1     &  AFM85         \\    
 10 & J203243.01+411529.9     &                \\    
 11 & J203241.84+411422.14    &  MT309         \\    
 12 & J203241.80+411439.2     &                \\    
 13 & J203240.89+411429.6     &  MT304         \\    
 14 & J203240.35+411420.1     &                \\    
 16 & J203239.90+411436.2     &                \\    
 17 & J203238.51+411541.2     &  MT297         \\    
 18 & J203238.09+411441.8     &                \\    
 19 & J203237.89+411509.7     &  MT294         \\    
 20 & J203236.58+411644.7     &                \\    
 21 & J203235.63+411509.6     &  AFM27         \\    
 22 & J203235.33+411445.3     &  MT282         \\    
 23 & J203233.44+411406.0     &                \\    
 24 & J203233.07+411359.6     &                \\    
 25 & J203231.49+411408.4     &  AFM17         \\    
\hline 
\end{tabular}
\end{table}
       Images of the region around MT304 in B and V filters have been acquired on May 30 2013 using the direct imaging mode of SCORPIO focal reducer  \citep{scorpio} mounted on the Russian 6-m telescope, and on October 14 2012 using the CCD photometer of Zeiss-1000 telescope of Special Astrophysical Obsery of Russian Academy of Science (SAO RAS). The observations are summarized in Table~\ref{tab:log}. The images have been bias subtracted and flat-fielded.  The field is not crowded and the stars are well separated and well sampled, so we performed the aperture photometry using the {\sc SExtractor} package \citep{sextractor}. We applied circular apertures with the radius equal to mean full width at half-maximum (FWHM) measured over the {\color{black} stellar profile}. The FWHM has been quite stable over the field in all filters we used. No aperture corrections are taken into account. Instead, we calibrated the instrumental magnitudes for every image by measuring the stars with B and V photometry published by \citet{MT91}.  To solve the photometric equation linking the instrumental magnitudes in two bands to catalogued values in standard photometric bandpasses, we iteratively computed using weighted least squares the mean zero point (which includes both the instrumental zero point and the atmospheric extinction) and the color correction term (proportional to the difference of instrumental magnitudes in two bands) and then excluded the calibration stars whose measurements deviate from \citet{MT91} values for more than two ensemble standard deviations. Then, using the derived parameters of photometric equation, we converted our instrumental magnitudes for every star to the magnitudes in photometric system of \citet{MT91}. 
       
       The resulting number of calibration stars used to solve the photometric equation on each image is listed in Table~\ref{tab:log}. The uncertainly of this zero point is the primary source of errors in our photometry, as the internal accuracy of flux measurements is typically much less than 0.01 magnitude, and it may therefore may be considered the error of our photometric measurements. It is also listed in Table~\ref{tab:log}.

       Most of the stars, except for MT340, MT304 and AFM17, are unsaturated on 30-s exposure images acquired with SCORPIO, so we used these frames to measure  their magnitudes. Photometry of the brightest stars has been performed on 1-second exposure frame in V filter and 3-second exposure frame in B filter from the same instrument. One star, USNO-B1.0 1312-0389914, is  outside the SCORPIO frames, and we measured it from images acquired by Zeiss-1000 telescope and using the same procedure.
       
       Table~\ref{tab:parameters} shows the photometry of all stars in our sample, as well as their B-V colors.
       
      \subsection{Speckle-interferometery}\label{sec:specl}

       \citet{Caballero} discovered a second component of MT304, with the {\color{black} separation} between the components of $63.6~{\rm mas}$ and the magnitude difference of $\Delta $m=$2.3\pm0.2$~mag {\color{black} in F583W filter}. To confirm it, and to measure the brightness difference on other spectral bands, as well as to  better study the immediate neighbourhood of MT304 in general, we performed speckle-interferometric observations with the Russian 6-m telescope on February 12 and December 5 2014. The speckle interferometer we used is based on the fast and low noise electron-multiplying CCD (EMCCD) with $512 \times 512$ elements \citep{maks2009}. It is equipped with $\times16$ microobjective providing 4.5\arcsec$\times$4.5\arcsec \ field of view with $9.118\,\pm 0.004$~milliarcsecond (mas) per pixel angular resolution, and it reaches the diffraction limit, which is $0.02$\arcsec \ for a 6-m telescope in the visual range ($\lambda\sim5000$~\AA) and $0.033$\arcsec \ in the near-infrared  ($\lambda\sim8000$~\AA). The observations run typically implies the acquisition of 2000 consecutive images with 20\,ms exposure, every image  -- speckle interferogram -- being the convolution of an instant point-spread function, defined mainly by the atmospheric turbulence, with the true brightness distribution of the studied object \citep{Labeyrie1970}.   

       Methods of speckle interferometric data reduction from 6-m telescope are described in \citet{maks2009} and \citet{bal2002}. For the point-like objects they are based on the analysis of the autocorrelation functions and two-dimensional power spectra computed for every image and then averaged over the sequence. Every companion present in the system generates two symmetric peaks in the autocorrelation, and the series of fringes in power spectrum, with fringe contrast and period defined by the magnitude difference and angular separation of the companion from the central object. Fringes orientation determines position angle of components with uncertainty $\pm 180\,\degr$. Due to high dynamic range of the EMCCD camera we use, magnitude difference up to 6\,mag can be measured. 

       We also used the image reconstruction method based on the bispectral analysis method \citep{Lohmann1983} which allows us to estimate the true brightness distribution of the system and to reliably determine the positional angle of the companions. 

       Observations were carried out in the visual and near-infrared spectral ranges with filters having central wavelengths of 7000, 8000 and 9000~\AA \ and pass-band halfwidths of 400, 1000 and 800 \AA \ correspondingly (Table~\ref{tab:specl}). Atmospheric seeings, estimated using the images averaged over the whole sequence (which is equivalent to an usual long-exposure imaging), were 1.2\arcsec \ and 1.5\arcsec, correspondingly. 

\begin{figure*}
\includegraphics[scale=0.4,viewport=80 80 678 655,clip]{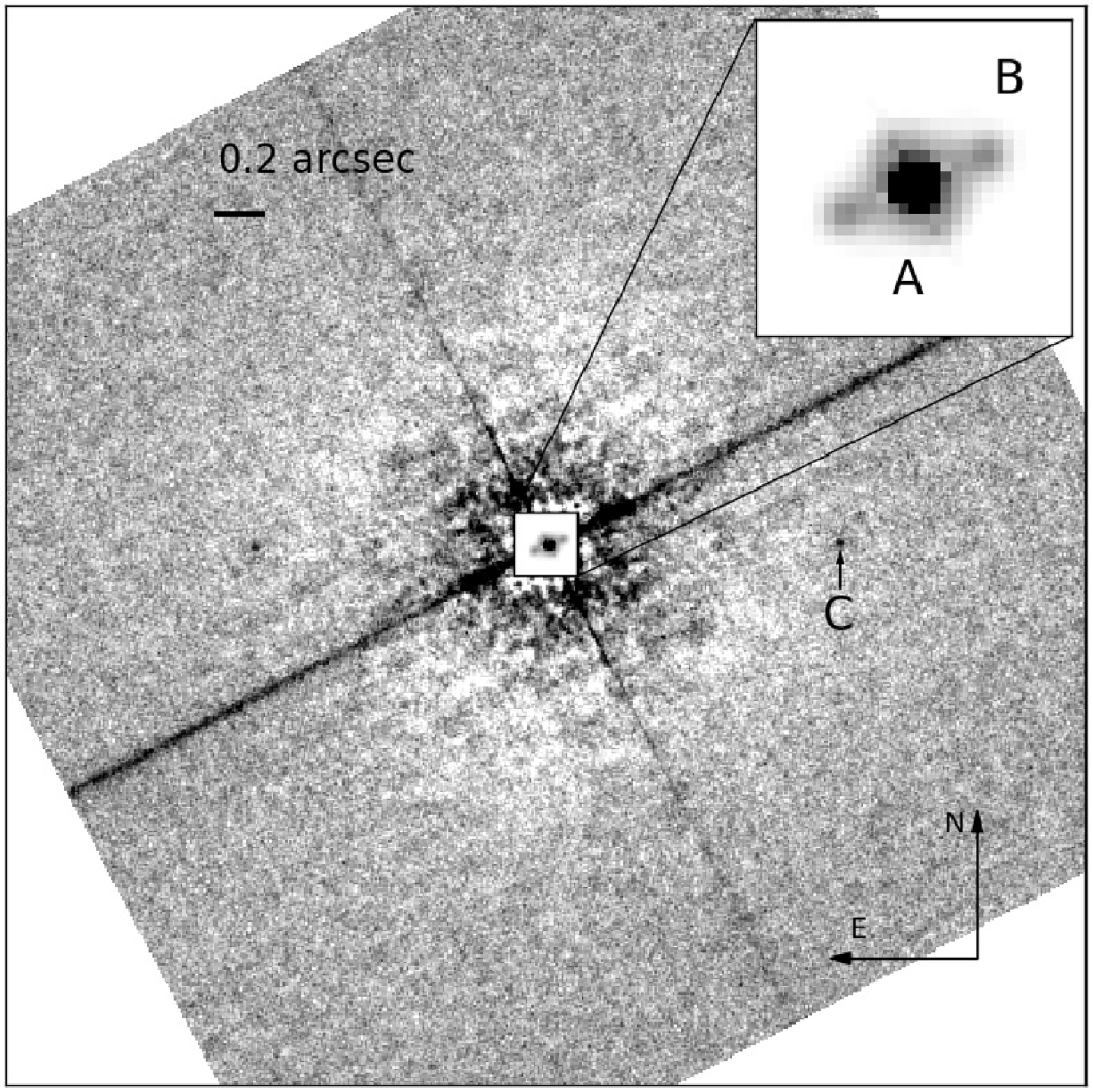}
\includegraphics[scale=0.4,viewport=80 80 678 655,clip]{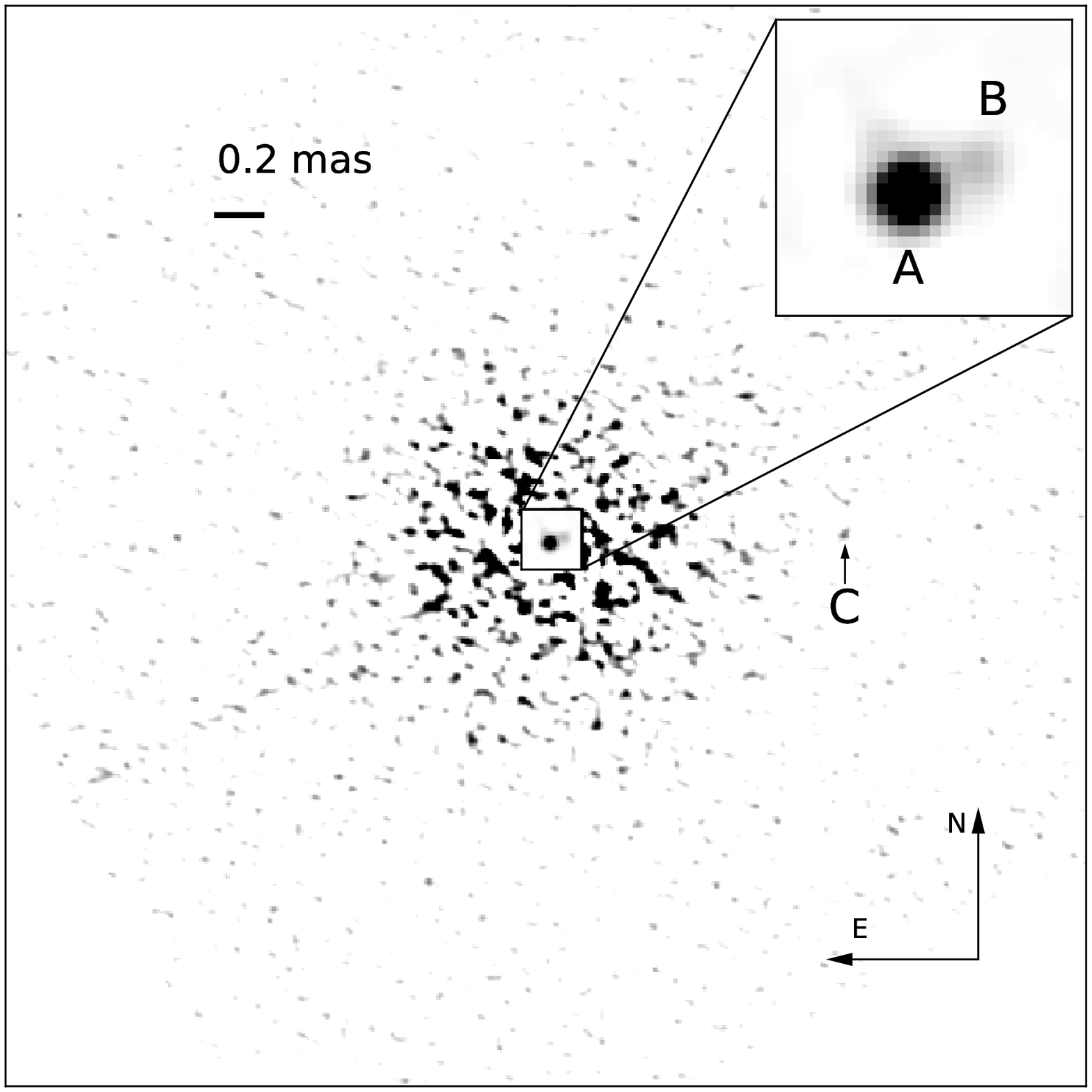} \\ 
\includegraphics[scale=0.4,viewport=80 80 678 655,clip]{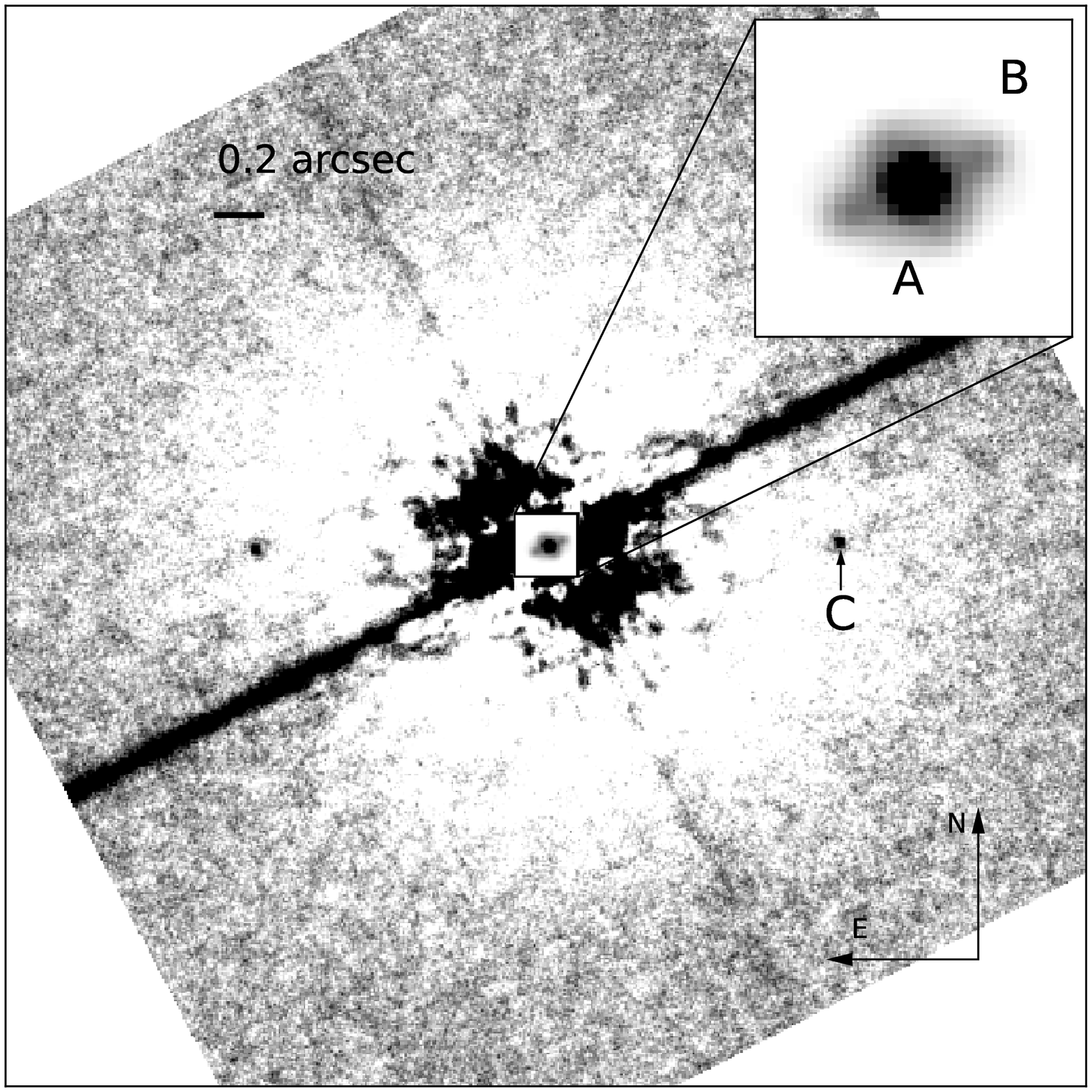}
\includegraphics[scale=0.4,viewport=80 80 678 655,clip]{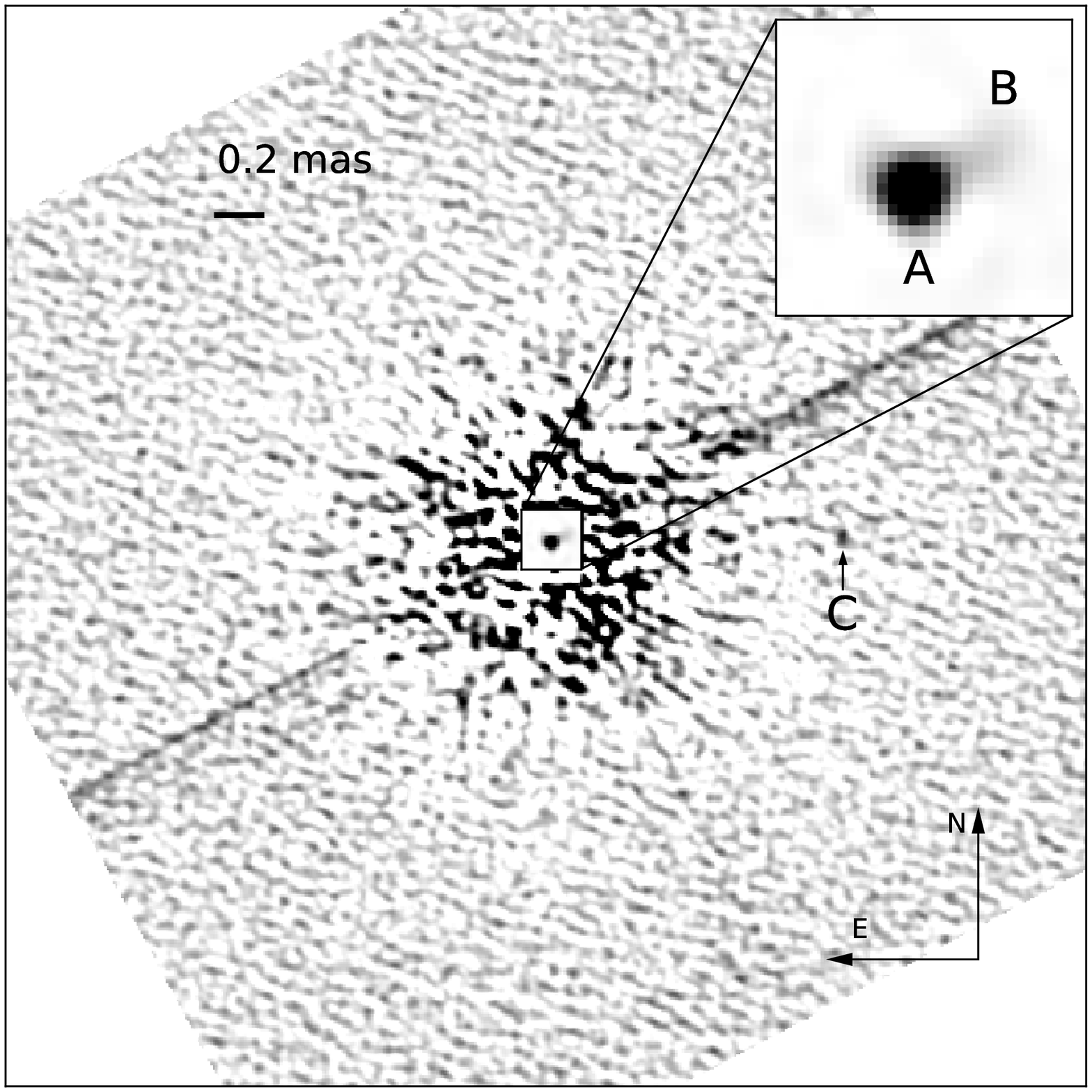}                          
\caption{The autocorrelation functions (left) and the reconstructed images (right) of MT304 multiple system in the 7000/400~\AA \ filter (upper panel) and  
         8000/1000~\AA \  (lower panel). Observations  were carried out on February 12 2014.}
\label{fig:acf}
\end{figure*}

\begin{figure*}
\includegraphics[scale=0.4,viewport=80 80 678 655,clip]{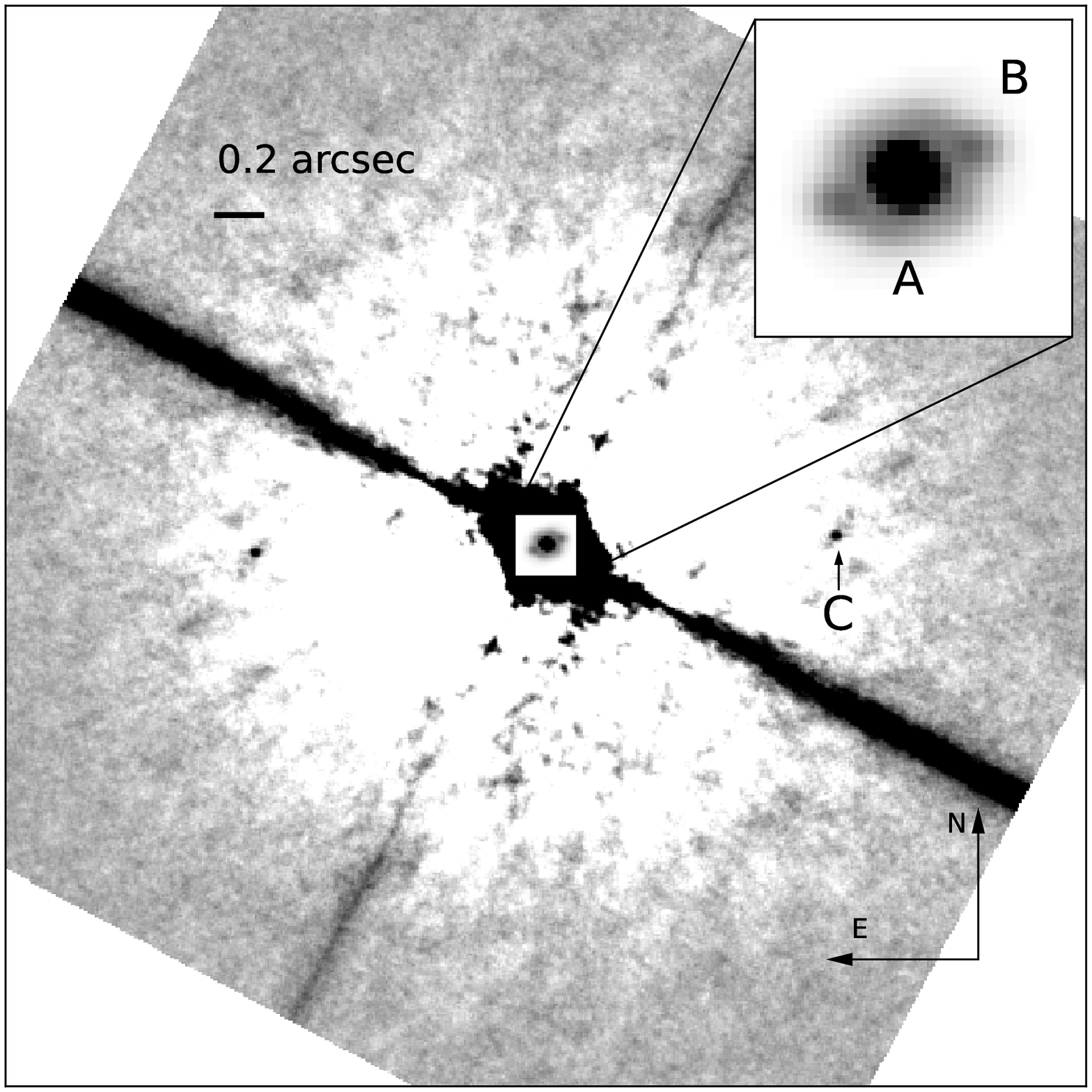}
\includegraphics[scale=0.4,viewport=80 80 678 655,clip]{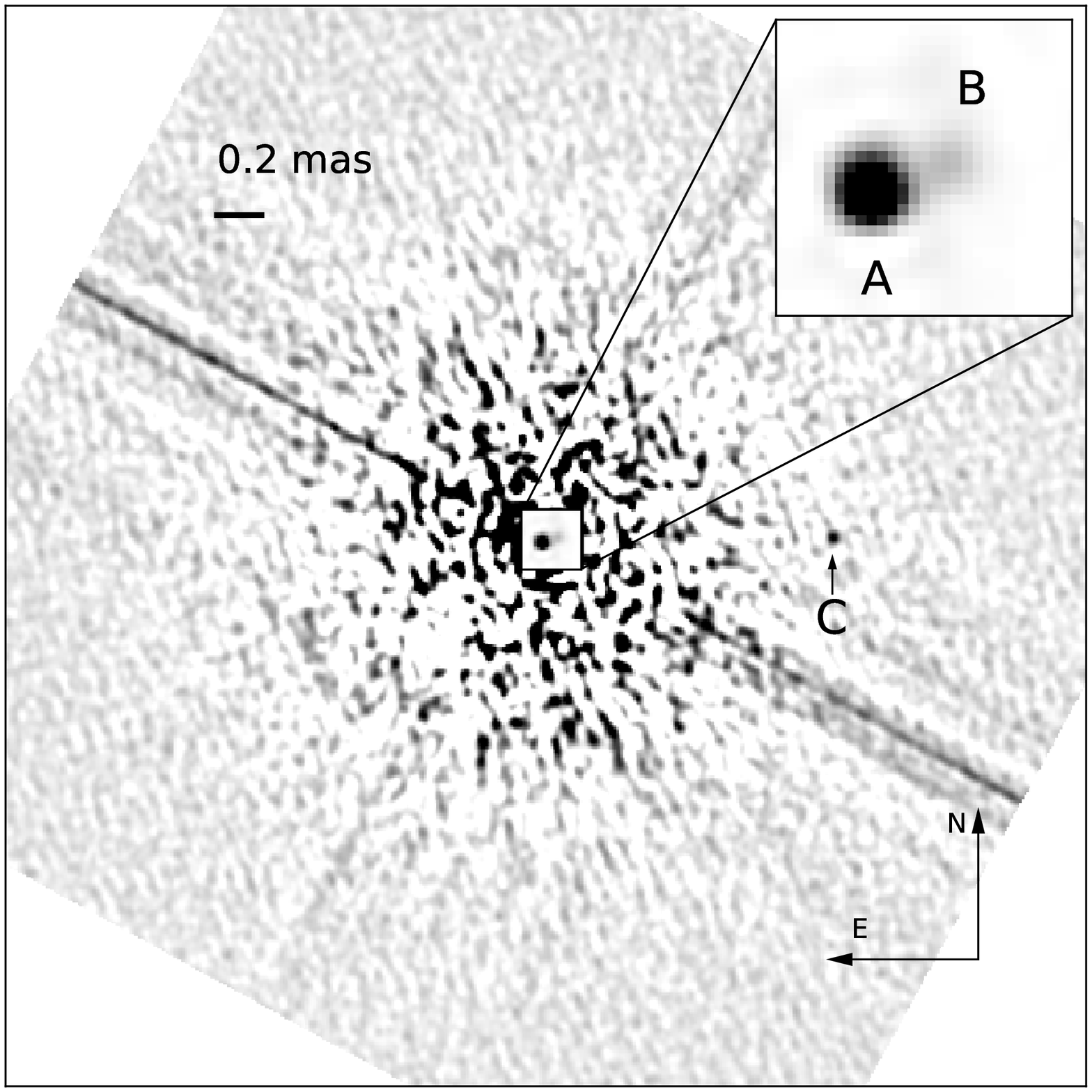} \\
\includegraphics[scale=0.4,viewport=80 80 678 655,clip]{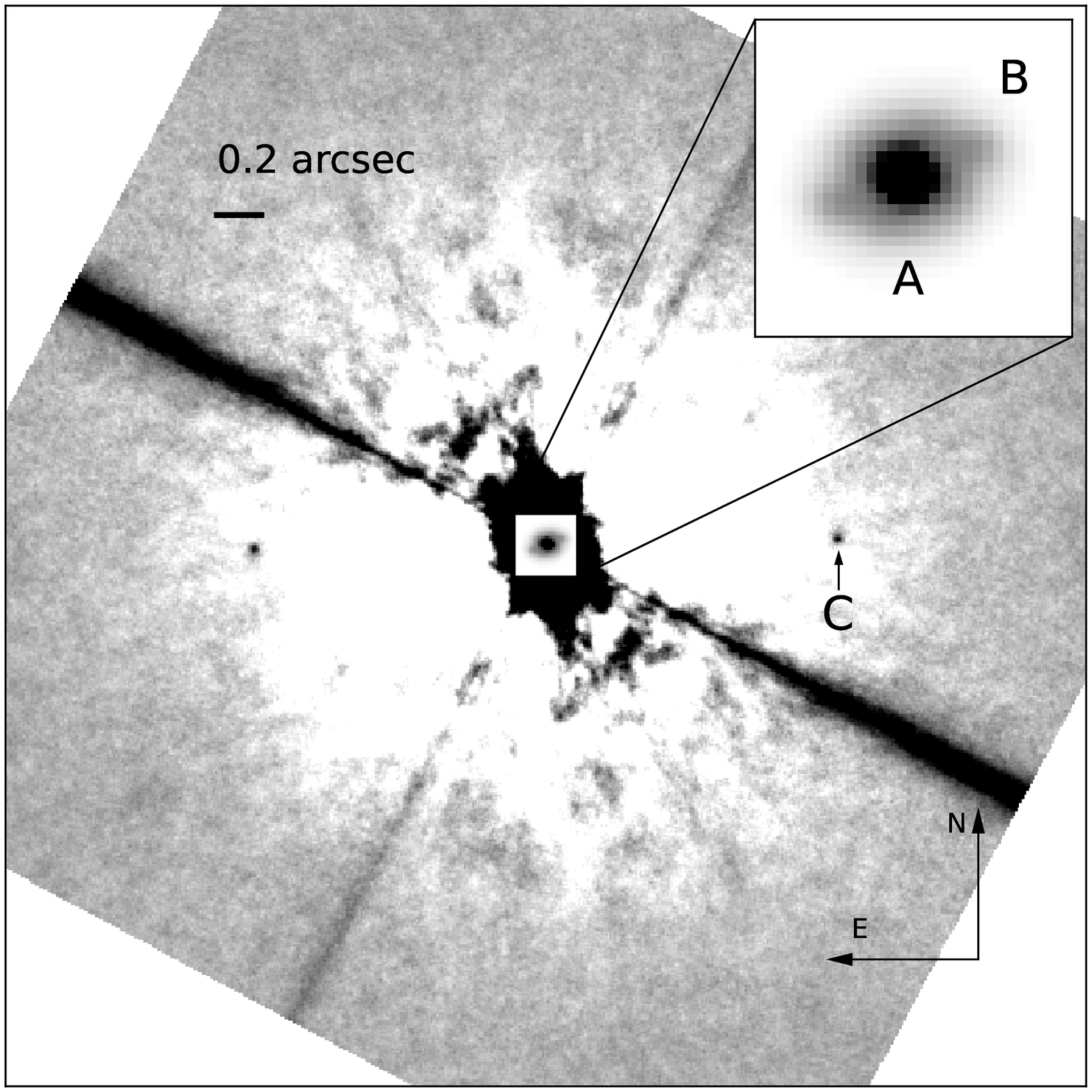}
\includegraphics[scale=0.4,viewport=80 80 678 655,clip]{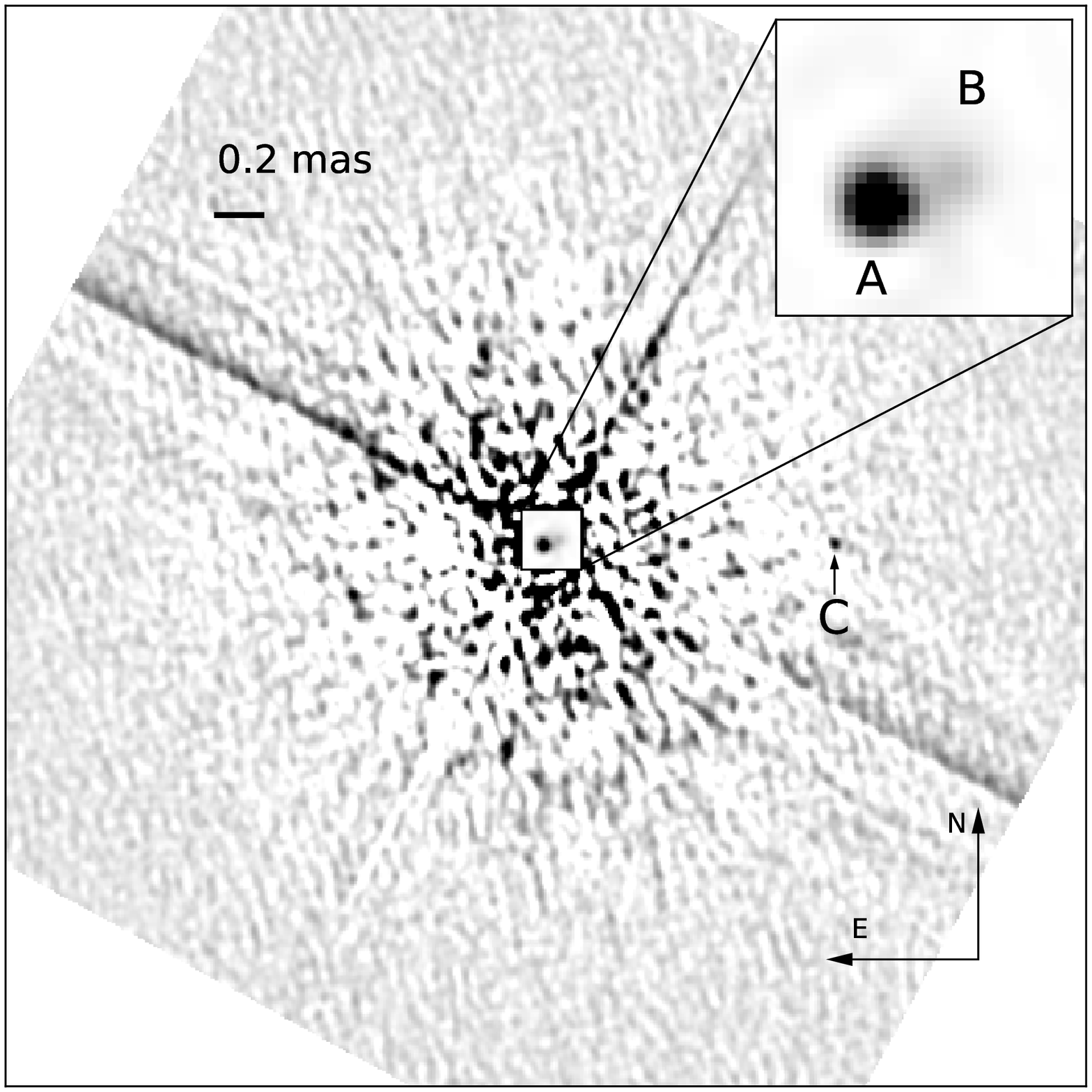}
\caption{The autocorrelation functions (left) and the reconstructed images (right) of MT304 multiple system in the 9000/800~\AA \ filter (upper panel) and  
         8000/1000~\AA \  (lower panel). Observations were carried out on  December 5 2014.}
\label{fig:acfdec}
\end{figure*}

       The autocorrelation function (see left panels of Figures~\ref{fig:acf} and~\ref{fig:acfdec}) clearly displays the second component of MT304, discovered by \citet{Caballero}. Moreover, the fainter third companion is also seen with the significance better than at least 7 sigmas. The reconstructed images, shown in right panels of  Figures~\ref{fig:acf} and \ref{fig:acfdec}, also display both components in all four sets of observations.  

       The measured parameters are given in Table~\ref{tab:specl}; for the second component they are well consistent with the estimates by \citet{Caballero}. 
 
\begin{landscape}
\begin{table}
\caption{Interstellar extinction and distance towards the stars close to MT304. First column is {\color{black} the source name in the catalog} by \citet{MT91} or \citet{AFM}, 
         2 -- name in Sloan Digital Sky Survey (SDSS) catalog. 
         Columns 3 and 4 point out the range of spectral subclasses, while columns 5, 6 and  7, 8 show the values of absolute star magnitude $M_V$ 
         and intrinsic color index (B-V)$_0$ corresponding to this range.  Following this logic, columns 14 and  15 point out the range of ${\rm A}_{V}$ 
         for all sample stars, while columns 16, 17 the range of distances. } 
\label{tab:parameters}
\vspace{0.5cm}
\begin{tabular}{l|l|ll|ll|ll|l|l|l|ll|ll|ll}
             &                        &\multicolumn{2}{c|}{~~~}       &\multicolumn{2}{c|}{$M_V$}  &\multicolumn{2}{c|}{(B-V)$_0$}& ~~B~ & ~~V~ & (B-V) & \multicolumn{2}{c|}{${\rm E}_{B-V}$} &  \multicolumn{2}{c|}{${\rm A}_{V}$}   & \multicolumn{2}{c}{D}    \\  
  \multicolumn{1}{c}{AFM/MT}     &\multicolumn{1}{c}{SDSS}&\multicolumn{2}{c|}{Spec. type}&\multicolumn{2}{c|}{[mag]}  &\multicolumn{2}{c|}{[mag]}    & [mag]& [mag]& [mag] & \multicolumn{2}{c|}{[mag]}           &  \multicolumn{2}{c|}{[mag]}           & \multicolumn{2}{c}{[kpc]}\\                
 \multicolumn{1}{c}{(1)}         &\multicolumn{1}{c}{(2)} &       \multicolumn{1}{c}{(3)}   &       \multicolumn{1}{c}{(4)}       &   \multicolumn{1}{c}{(5)}       &        \multicolumn{1}{c}{(6)}   &    \multicolumn{1}{c}{(7)}       &       \multicolumn{1}{c}{(8)}     &  \multicolumn{1}{c}{(9)} &  \multicolumn{1}{c}{(10)}&  \multicolumn{1}{c}{(11)} &    \multicolumn{1}{c}{(12) }             &     \multicolumn{1}{c}{(13)}      &     \multicolumn{1}{c}{(14)}      &     \multicolumn{1}{c}{(15)}              &         \multicolumn{1}{c}{(16)}     &   \multicolumn{1}{c}{(17)}\\
\hline             
             \multicolumn{16}{l}{Stars between us and the association}\\
   MT340     & J203250.25+411448.4   &      G8 V   &    G9 V         & +5.5        &     +5.7     &    +0.74     &     +0.775    & 14.68 &13.7  & 0.98 &       0.24       &       0.205       &     0.72          &        0.615      &    0.31      & 0.3        \\    
             & J203247.28+411339.0   &      K0 V   &    K7 V         & +5.9        &     +8.1     &    +0.81     &     +1.33     & 18.53 &17.08 & 1.45 &       0.64       &       0.13        &     1.94          &        0.38       &    0.7       & 0.5        \\    
             & J203243.84+411446.5   &      K0 V   &    K2 V         & +5.9        &     +6.4     &    +0.81     &     +0.91     & 20.9  &19.2  & 1.7  &       0.9        &       0.8         &     2.7           &        2.4        &    1.3       & 1.2        \\    
   AFM85     & J203243.49+411445.1   &      K7 V   &    M2 V         & +8.1        &     +9.9     &    +1.33     &     +1.49     & 19.63 &18.05 & 1.6  &       0.27       &        0.1        &     0.8           &        0.3        &    0.7       & 0.4        \\    
             & J203243.01+411529.9   &      M2 V   &    M5 V         & +9.9        &     +12.3    &    +1.49     &     +1.64     & 19.37 &17.73 & 1.64 &       0.15       &       0           &     0.45          &        0          &    0.3       & 0.1        \\    
   MT309     & J203241.84+411422.14  &      G8 V   &    K1 V         & +5.5        &     +6.15    &    +0.74     &     +0.86     & 16.0  &15.0  & 1.00 &       0.26       &       0.13        &     0.8           &        0.4        &    0.6       & 0.5        \\    
   MT297     & J203238.51+411541.2   &      G8 V   &    K0 V         & +5.5        &     +5.9     &    +0.74     &     +0.81     & 17.3  &16.14 & 1.16 &       0.42       &        0.35       &     1.16          &        1.25       &    0.75      & 0.7        \\    
   MT294     & J203237.89+411509.7   &      F0 V   &    F1 V         & +2.7        &     +3.15    &    +0.3      &     +0.325    & 16.64 &15.41 & 1.23 &       0.93       &        0.9        &     2.79          &        2.71       &    1.0       & 0.8        \\    
             & J203236.58+411644.7   &      K1 V   &    K7 V         & +6.15       &     +8.1     &    +0.86     &     +1.33     & 22.4  &20.44 & 1.96 &       1.1        &       0.64        &     3.3           &        1.9        &    1.6       & 1.2        \\    
             & J203233.07+411359.6   &      F8 V   &    G2 V         & +4.0        &     +4.7     &    +0.52     &     +0.63     & 18.93 &17.43 & 1.5  &       0.98       &       0.87        &     2.94          &        2.6        &    1.25      & 1.05       \\    
             &                       &             &                 &             &              &              &               &       &      &      &                  &                   &                   &                   &              &            \\
\cline{1-2} 
\multicolumn{16}{l}{Stars  member of the association}\\
  MT343     & J203250.75+411502.2   &      B1 V   &    B2 V         & -3.2        &     -2.45    &    -0.26     &     -0.24     & 16.4   &14.41  & 2.00 &       2.26       &       2.24        &     6.78          &        6.72        &    1.5       & 1.1      \\    
  MT333     & J203248.62+411429.8   &      B1 V   &    B2 V         & -3.2        &     -2.45    &    -0.26     &     -0.24     & 17.9   &15.6   & 2.30 &       2.56       &       2.54        &     7.7           &        7.63        &    1.6       & 1.2       \\    
  AFM91     & J203245.07+411416.5   &      B8 V   &    A1 V         & -0.25       &     +1.0     &    +0.11     &     +0.01     & 21.2   &18.93  & 2.27 &       2.38       &        2.27       &     7.16          &        6.8         &    2.5       & 1.6       \\    
  AFM79     & J203243.61+411513.9   &      B5 V   &    B7 V         & -1.2        &     -0.6     &    -0.17     &     -0.13     & 19.85  &17.62  & 2.23 &       2.4        &        2.36       &     7.2           &        7.0         &    2.0       & 1.6       \\    
            & J203241.80+411439.2   &      B5 V   &    B9 V         & -1.2        &     +0.2     &    -0.17     &     -0.07     & 21.8   &19.05  & 2.75 &       2.93       &       2.83        &     8.78          &        8.5         &    1.9       & 1.2       \\    
  MT304     & J203240.89+411429.6   &      B5 Ia-0&                 & -8.4        &              &    -0.08     &               & 14.76  &11.5   & 3.26 &       3.34       &                   &     10.0          &                    &    1.1       &           \\    
            & J203240.35+411420.1   &      B2 V   &    B5 V         & -2.45       &     -1.2     &    -0.24     &     -0.17     & 20.67  &17.86  & 2.81 &       3.05       &       2.98        &     9.15          &        8.94        &    1.7       & 1.1       \\    
            & J203239.90+411436.2   &      B5 V   &    B6 V         & -1.2        &     +0.9     &    -0.17     &     -0.15     & 21.68  &18.84  & 2.84 &       3.01       &       3.0         &     9.03          &        9.00        &    1.6       & 1.4      \\    
            & J203238.09+411441.8   &      B5 V   &    B7 V         & -1.2        &     -0.6     &    -0.17     &     -0.13     & 20.93  &18.33  & 2.6  &       2.78       &       2.74        &     8.32           &        8.2         &    1.7       & 0.8       \\    
  AFM27     & J203235.63+411509.6   &      B1 V   &    B2 V         & -3.2        &     -2.45    &    -0.26     &     -0.24     & 18.2   &15.9   & 2.32 &       2.58       &        2.56       &     7.75          &        7.7         &    1.8       & 1.4       \\    
  MT282     & J203235.33+411445.3   &      B1 V   &    B1 IV        & -3.2        &     -3.8     &    -0.26     &               & 17.31  &14.9   & 2.42 &       2.68       &                   &     8.04          &                    &    1.0       & 1.4       \\    
  AFM17     & J203231.49+411408.4   &      O7.5 Ib&    IIf          & -6.3        &     -5.8     &    -0.31     &     -0.32     & 15.14  &12.9   & 2.24 &       2.56       &    2.57           &     7.68           &        7.7        &    2.0       & 1.6       \\    
            &USNO-B1.0 1312-0389914 &      B7 V   &    A2 V         & -0.6        &     +1.3     &    -0.13     &     +0.05     & 19.19  &17.27  & 1.92 &       2.05       &    1.87           &     6.15          &        5.61        &    2.2       & 1.2       \\    
            &                       &             &                 &             &              &              &               &        &       &      &                  &                   &                   &                    &              &           \\    
\cline{1-2}
\multicolumn{16}{l}{Stars behind the association}\\
           & J203233.44+411406.0   &      B5 V   &    B7 V         & -1.2        &     -0.6     &    -0.17     &     -0.13     & 20.53  &18.31   & 2.22 &       2.39       &       2.35        &     7.16          &        7.04       &    2.9       & 2.3        \\    
           & J203247.17+411501.0   &      G8 III &    K0 III       & +0.8        &     +0.7     &    +0.94     &     +1.0      & 21.73  &18.78   & 2.95 &       2.01       &       1.95        &     6.02          &        5.85        &    2.5       & 2.8     \\    
           &                       &             &                 &             &              &              &               &        &        &      &                  &                   &                   &                   &              &            \\    
\cline{1-2}
  \multicolumn{16}{l}{* -- values of  $M_V$ and (B-V)$_0$ are taken from \citet{SchmidtKaler}.} 
\end{tabular}
\end{table}
\end{landscape}

\begin{table*}
\caption{Resolved companions of MT304. $\theta$ is the measured position angle, $\rho$ is the measured angular separation, $\Delta$m is the observed magnitude difference, $\lambda$ is the central wavelength of the filter used for the observation, $\Delta\lambda$ -- the full width at half-maximum (FWHM) of the filter passband. }
\label{tab:specl}
\begin{tabular}{lcccccc}
\hline
      Pair   &   Epoch   & $\theta$, [$\degr$]         & $\rho$         & $\Delta$m      & $\lambda$ & $\Delta\lambda$ \\
             &           &                             & [mas]          & [mag]          & [nm]      &  [nm]          \\
\hline
AB           & 2014.1198 & $293.0\pm 0.3$              & $65\pm   1$    & $1.79\pm 0.02$ & 700       & 40             \\
             & 2014.1199 & $293.0\pm 0.3$              & $64\pm   2$    & $1.75\pm 0.03$ & 800       & 100            \\
             & 2014.9290 & $291.7\pm 0.3$              & $64\pm   2$    & $2.0 \pm 0.1 $ & 800       & 100            \\
             & 2014.9290 & $291.7\pm 0.3$              & $69\pm   2$    & $1.8 \pm 0.1 $ & 900       & 80             \\
\\
AC           & 2014.1198 & $271.3\pm 0.2$              & $1246\pm 2$    & $4.8 \pm 0.2 $ & 700       & 40             \\
\hline
\\
AB$^*$       & 2006.2346 & ${305.9\atop283.5} \pm 3.3$ & $63.5\pm3.5$   & $2.3\pm0.2$  & 583       & 234            \\ 
\\
\multicolumn{7}{l}{$^*$ Data taken from \citet{Caballero} }\\
\hline
\end{tabular}
\end{table*}

\section{Results}\label{sec:res}

         We performed the spectral classification of the stars in our sample in two different ways: 1) qualitatively using spectral atlases by \citet{Jacoby} and \citet{WalbornFitzpatrick90}, and 2) automatically by means of $\chi^2$ fitting with spectral standards from STELIB\footnote{STELIB http://webast.ast.obs-mip.fr/stelib} (see \citet{STELIB}). In the latter method, similar to the one used by \citet{AbolmasovW50}, we normalized the spectrum of studied star to the continuum level, and then compared it with the spectra taken from STELIB, convolved with spectral resolution of the observed spectrum.

         For nearly all sample stars the spectral types estimated with both methods are the same; when they differ we widen the uncertainty of the {\color{black} spectral type} in  the Table~\ref{tab:parameters} to cover the results of both methods.  For determination of  {\color{black} the luminosity class of the  B stars} we used hydrogen lines which weaken with increasing luminosity (see, for example, \citet{StellarSpectralClassification}). For later spectral {\color{black}  classes} we also used ratios of luminosity sensitive lines. {\color{black} We assume that those stars  for which we could not estimate the luminosity class are dwarfs, i.e. luminosity class V.} This assumption is based on the much greater number of dwarf stars. Anyway, it does not introduce significant errors for the determination of reddening, as the dwarfs and giants has similar intrinsic color indices.
         By combining estimates of spectral types with the photometric data we measured the individual interstellar extinction ${\rm A}_{V}$:
          $${\rm A}_{V}={\rm R}_{V}\cdot{\rm E}_{B-V}$$
         where ${\rm E}_{B-V}=(B-V)-(B-V)_0$ and we assume ${\rm R}_{V}=3$ according to   \citet{Hanson2003}.  For estimation of distances D towards the stars we  used the classical formula:
         $$          log~D=0.2\times(V-M_V-{\rm A}_{V}+5).         $$
         Fundamental stellar parameters -- such as absolute star  magnitudes $M_V$ and intrinsic color indices (B-V)$_0$ -- were taken  from \citet{SchmidtKaler}. The range of possible spectral types for every star corresponds to the range of $M_V$ and (B-V)$_0$, which in turn give the range of interstellar extinction and distances.  

         Together with  MT304, only MT343 (J203250.75+411502.2)  and  J203231.49+411408.4  were previously studied spectroscopically. Our spectral classification of MT343 as B1~V  star is consistent with  \citet{KiminkiAv}. For J203231.49+411408.4\footnote{Object J203231.49+411408.4 was included in the list of members of Cyg~OB2  association  by \citet{Comeron02} based on near-infrared spectroscopy and JHK-photometry and named  A11. The object was identified with X-rays source  AFM17  by \citet{AFM}. We, as \citet{Kobulnicky}, suppose that J203231.49+411408.4 is the same object as MT267, although \citet{KiminkiAv} classified MT267 as G-type star.  Interstellar extinction $A_V$ measured by (B-V)  agrees with one measured by \citet{Comeron12} using 2MASS $(J-K_S)$,  and the distance we estimated is an evidence that  J203231.49+411408.4 belongs to the Cyg~OB2.}
         we adopt the classification of   \citet{Negueruela} as a O7.5 Ib-II(f) star because the spectrum of this star we obtained is of low resolution.

         We can not reliably determine the luminosity class of  J203247.17+411501.0, which has  G8-K0 spectral type. The interstellar extinction toward this star, $A_V\simeq6.1$ mag, is fairly large. If we assume that the star is a dwarf, using the fundamental stellar parameters $M_V$ and (B-V)$_0$ for G8-K0~V type, we find that the star is at a distance of about 200~pc, and if it is subgiant -- the distance increases to 700-800~pc. But such high interstellar extinction  is not typical for the stars in solar neighborhood. According to the study  of \citet{Sale2009}, for the stars in the field of Cyg~OB2 interstellar extinction reaches 5~mag at a distance of more than 2~\kpc. Therefore, we assume that J203247.17+411501.0 is a giant and it is located behind the association at about 2.5~\kpc.

         The star J203235.33+411445.3 or MT282 has a spectrum  typical of B1-type and strong ($A_V>8$~mag) interstellar extinction, therefore we decide that the object lies in the association and classify it as B1~IV. The fact that the absorption lines in the spectrum of MT282 are not as deep as those in the spectrum of MT343 (classified as B1~V dwarf) supports the classification of MT282 as a subgiant.
         
         Table~\ref{tab:parameters} shows that about half of the stars lie between us and the  Cyg~OB2 association. Interstellar  extinction 
         clearly increases with distance and also becomes inhomogeneous on  $d>1$~kpc. It is basically consistent with interstellar dust reddening map of \citet{dustmap}.
         Moreover, there is one star J203233.44+411406.0 situated behind the association like J203247.17+411501.0. Figure~\ref{fig:map1} shows stars lying in front of, in and behind the association by different symbols.

\section{Discussion}\label{sec:disc}

\subsection{The secondary component of MT304}

    We conducted speckle interferometric observations of MT304 in February and December 2014. As can be seen in Table~\ref{tab:specl} during these 10 months the position angle of the secondary component component has changed by 1.7~$\deg$, greatly exceeding the measurement errors. It strongly favours the physical connection between A and B components. Our measurements together with the results of \citet{Caballero} suggest a rotation period of the secondary component of about 100-200 years. The motion of the secondary component is large enough to build the orbit of the system in ten years of observations and therefore to derive the mass ratio of these stars.
    
    The observed spectrum of MT304 does not display any lines of the secondary component (see e.g. \citet{Chentsov}) and is perfectly fit by a single star model \citep{Clark}, and therefore we may suggest that the secondary component is a B dwarf with spectrum similar to the one of  the brighter star. Dedicated observations with high signal to noise ratio are definitely necessary to classify it better. We can not presently say anything about the third component, but its spatial proximity suggests that it may belong to the Cyg OB2 association and be physically connected to MT304 too. We will perform further observations of this system.

\subsection{The stellar cluster around MT304}
 
\begin{figure}
\includegraphics[scale=0.42]{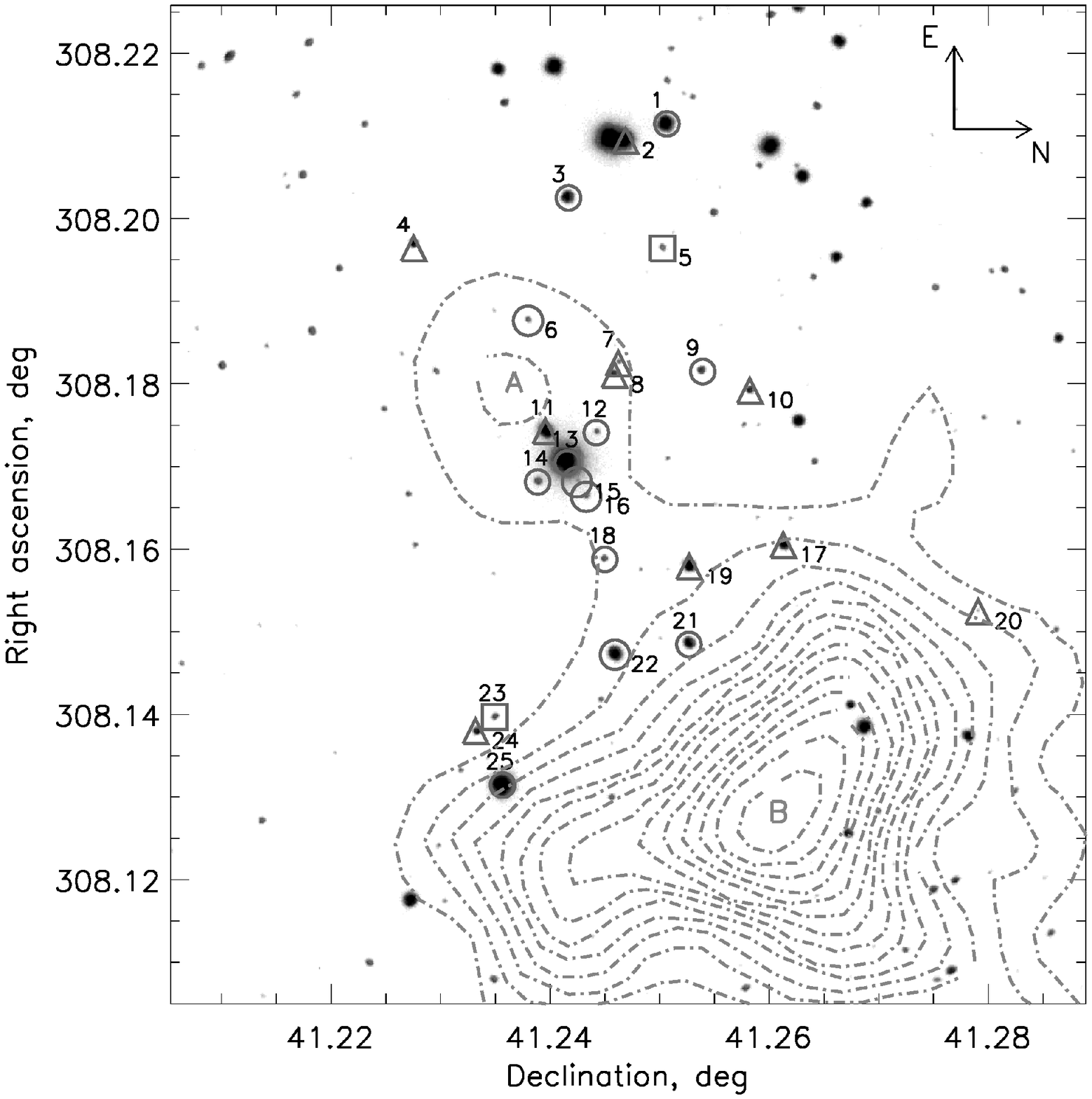}
\caption{R band image of the central part of Cyg~OB2 close to MT304 obtained with SCORPIO. Circles mark the stars belonging to the association, 
         triangles foreground stars and squares stars behind the association. Dot-dashed lines show contours of the $^{13}$~CO integrated 
         intensity $\int T_A^*~dv$. The contours range from 0.2 to 3.6~K~$\kms$ in steps of 0.3~K~$\kms$. The contours are taken from \citet{Scappini2002}.}
\label{fig:map1}
\end{figure}

     Direct imaging and speckle interferometric observations clearly display the cluster of stars around MT304. Figure~\ref{fig:map1} shows that MT304 is surrounded by five stars. MT309 (J203241.84+411422.14, marked by number 11 in the Figure~\ref{fig:map1}) is a foreground star according to our estimate of the distance (see Table~\ref{tab:parameters}). The stars marked as 12, 14 and 16 belong to Cyg~OB2 and they are the most reddened massive stars in Cyg~OB2 after MT304. One more star is marked by number 15. It is weak, with stellar magnitude greater than V=20 mag. We could not obtain its spectrum. Until we get new data we {\color{black} can not confirm the membership of this star.} We are planing future observations of this object. 
      
     Thus the multiple system MT304 is part of a group which consists of seven (three speckle-resolved components of MT304, three nearby hot stars definitely belonging to the association and one more star also supposedly belonging to it). 
     
\begin{figure}
\includegraphics[scale=0.7]{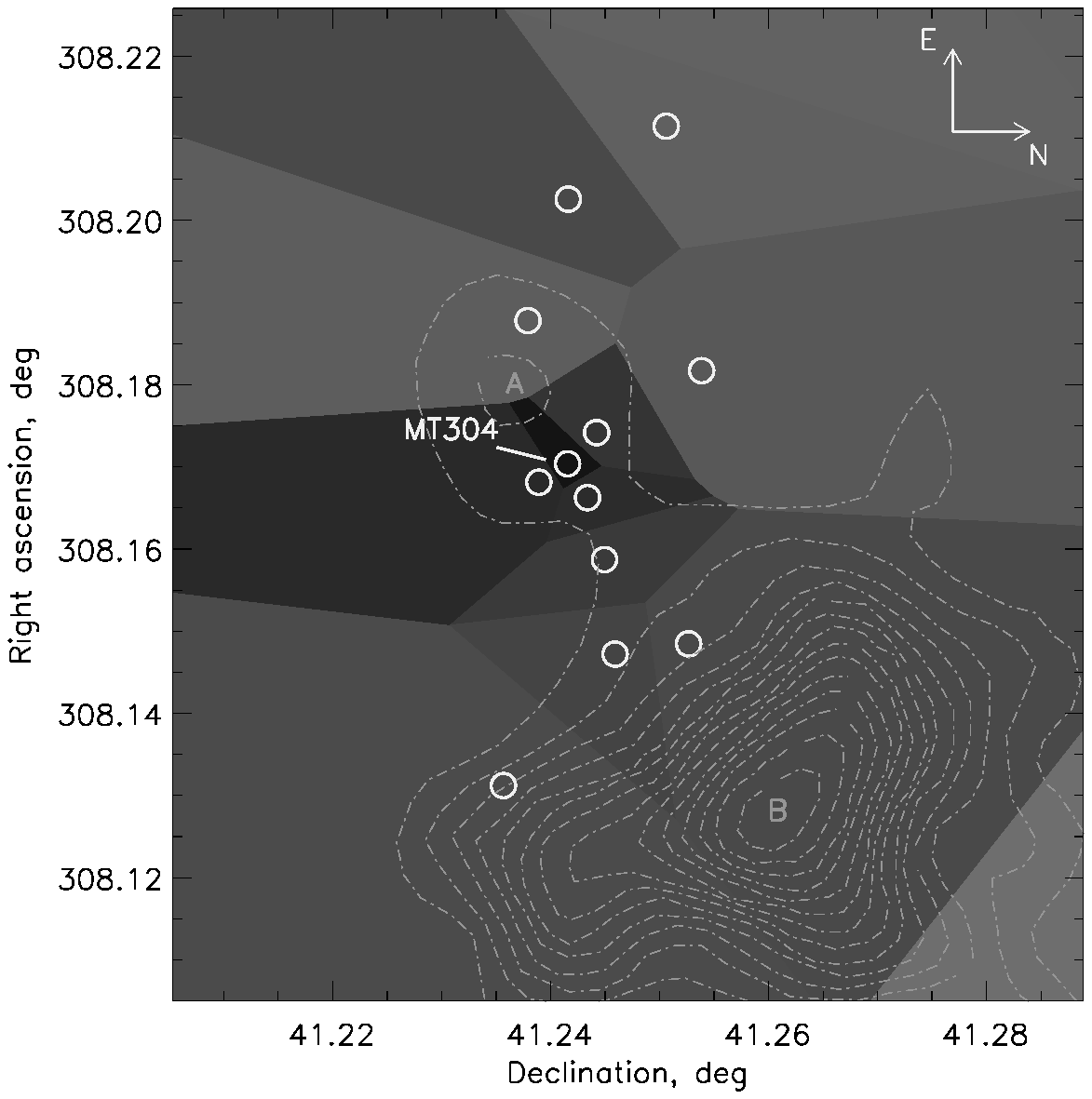}
\caption{Voronoi tessellated extinction map (one star per Voronoi cell) for twelve sample stars (except for USNO-B1.0 1312-0389914) 
         belonging to Cyg~OB2. As well as on the previous figure dash-dotted lines  show contours of the $^{13}$~CO integrated intensity $\int T_A^*~dv$.}
\label{fig:map2}
\end{figure}

\subsection{The nature of the reddening affecting MT304}    

%

         In this section we analyze the nature of the intense reddening affecting MT304, compared with the other massive stars of  Cyg~OB2.
         The typical $(B-V)_0$ color index for B5~Ia star is -0.08 \citep{SchmidtKaler}, while for MT304 it is 3.26. Hence the color excess is $E_{B-V}=3.34$~mag, that corresponds to $A_V=10$~mag, assuming ${\rm R}_{V}=3$ according to \citet{Hanson2003}. This excess can not  be  explained by an error in the spectral classification  (to reduce the extinction down to e.g. $A_V=9.0$~mag the star should be of F0 class, see \citet{SchmidtKaler})  or the distance to the star (which definitely belongs to the association, see  \citet{Chentsov} and discussion therein). It might be due to the fact that MT304 is a multiple system, and therefore its $(B-V)_0$ might differ from the one  of a single B5~Ia-0 star. However, the spectral lines of the second component are not detected  in the spectrum of  MT304, and therefore the second component should also be a B spectral type star, which can not significantly bias the spectral slope and therefore the extinction estimation. Contribution of the third companion may also be neglected due to large difference in brightness (nearly 5 mags). 
         Thus we may assume that the $A_V$ excess observed in  MT304  with respect to others stars is real.

         Table~\ref{tab:extinction} shows the comparison of our results with previous studies. Its upper half contains the most reddened OB stars with $A_V>7$ according to \citet{Wright2015},  while the lower part the most reddened objects as measured by us, selected from Table~\ref{tab:parameters}. Two stars, MT304 and AFM17, included in both parts of the table are shown in bold.   We analyzed the stars with V=13-20~mag and we increased the number of highly reddened massive stars, reducing the difference in reddening between MT304 and other members of the association. Before our study MT488 (USNO-B1.0 1312-0390173) was the second reddened massive star ($\Delta A_V=1.8$~mag) after MT304. Now J203240.35+411420.1 and J203239.90+411436.2 are the most reddened  massive stars and the difference is reduced down to $0.9\pm0.1$~mag. However MT304 remains the most reddened among massive stars in Cyg~OB2.
                  
\begin{figure*}
\includegraphics[scale=0.4,viewport=40 20 530 530,clip]{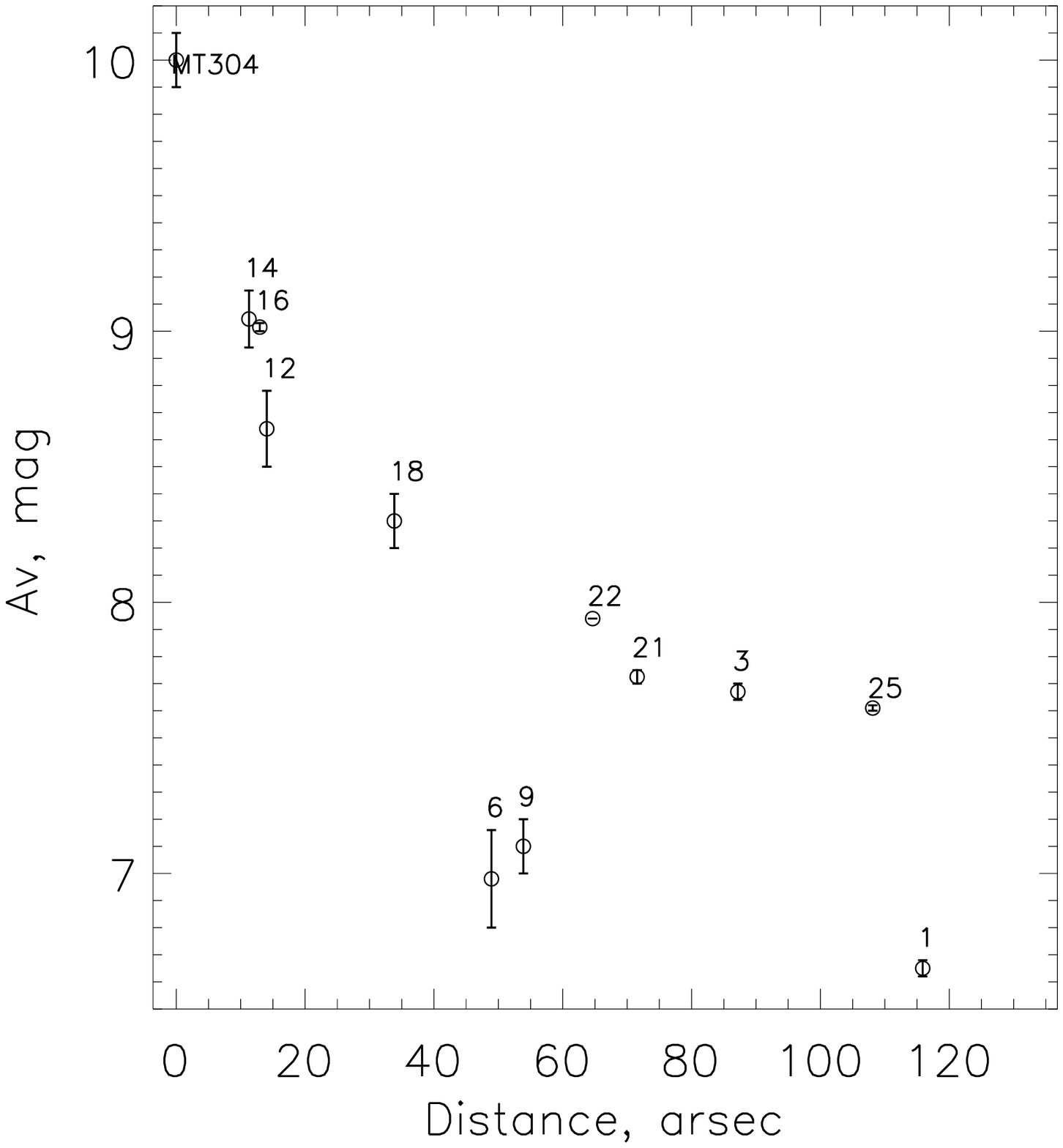}
\includegraphics[scale=0.4,viewport=40 20 530 530,clip]{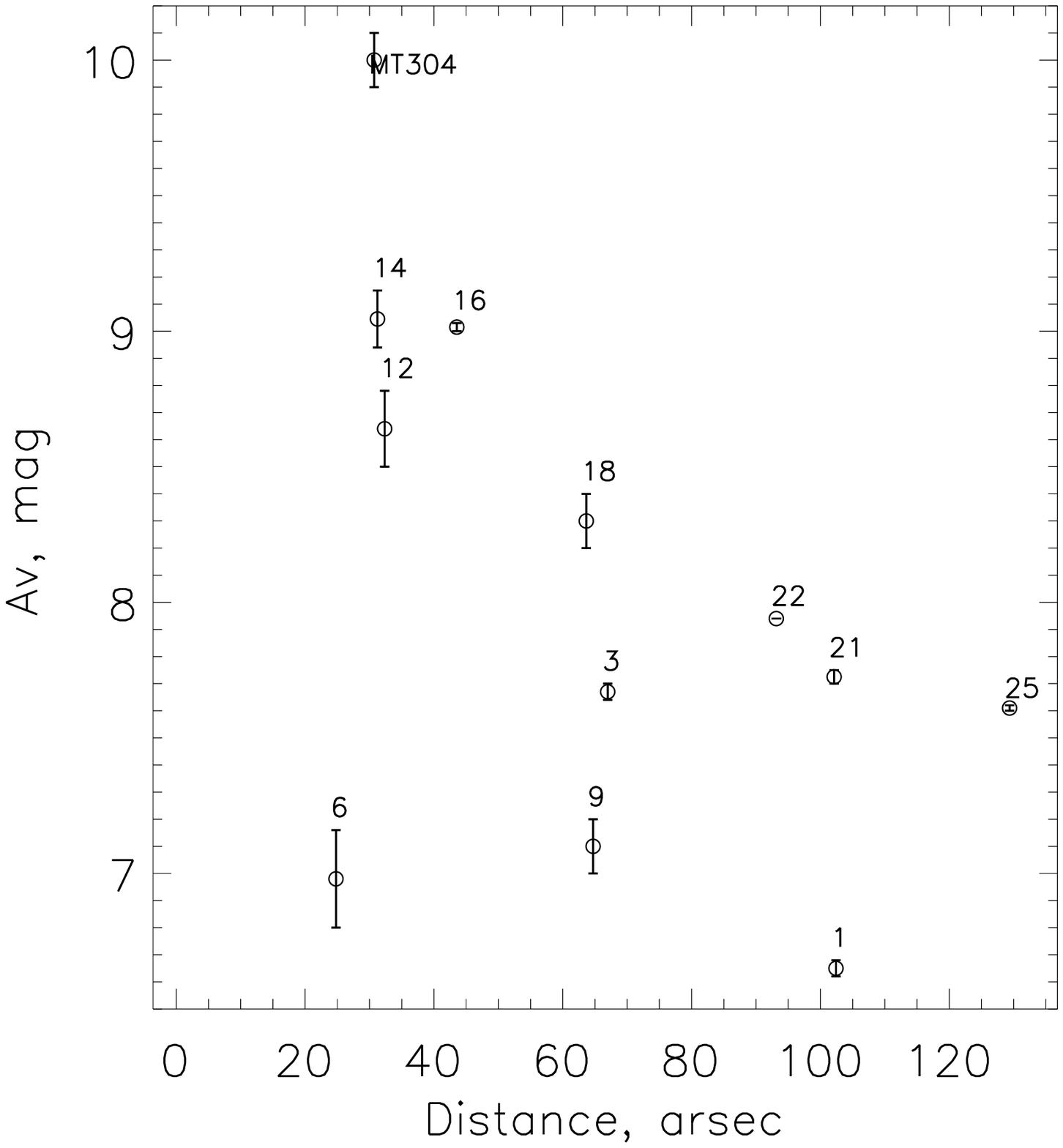}
\caption{Dependency of the interstellar extinction  $A_V$ on the distance to MT304 (left part) and on the distance to core A of the clump (right part).}
\label{fig:distav}
\end{figure*}

         Figure~\ref{fig:map2} shows the map of interstellar extinction based on the data shown in Table~\ref{tab:parameters}.  This figure is analogous to Figure~4 in \citet{Wright2015}, but with much better spatial resolution around MT304.  Extinction clearly increases with approaching to MT304, as it is shown in left panel of Figure~\ref{fig:distav}.  For the stars located within 30~arcsec from MT304 the extinction is higher than 8.5 mag (Table~\ref{tab:parameters}), i.~e. one magnitude greater than for  the other stars in this part of the association. 
         
         \citet{Whittet2015} explained the reddening toward MT304  as resulting from the accidental superposition of two or more clumps along the line of sight. \citet{Scappini2002} provided evidence for the existence of two molecular regions in its vicinity at 7 and 12~$\kms$.  Figure~\ref{fig:map2} displays the contours of the $^{13}$CO integrated intensity $\int T_A^*~dv$  at 12~$\kms$ taken from \citet{Scappini2002}, which shows the clump  consisting of two cores -- A and B (marked on  Figure~\ref{fig:map2}), core A being the nearest to MT304. Figure~\ref{fig:distav}, however, demonstrates that the $A_V$ quantity clearly depends on the distance from MT304, and does not show similar dependency on the distance from Core A  center. Besides, the stars 6, 12 and 14 in Figure~\ref{fig:map1} are at the same distance from the Core~A center as MT304, but their reddening is significantly lower. 
         Therefore, we suggest that the reddening excess is not directly related to the nearby clump. The same is also valid for for the second cloud, Core~B -- the stars 21, 22 and 25 in Figure~\ref{fig:map1} are closer to its center than MT304, but still have lower reddening ($A_V<8$~mag).
         
         
        {\color{black} It is unlikely that MT304 is still in its protostellar phase, whose lifetime is very short for such massive stars. We can exclude then the hypothesis that the high reddening affecting this stars is due to some surrounding collapsing parental cloud.} 
        It may be, however, attributed to the nebula ejected by MT304 itself during its evolution.  Circumstellar nebulae ejected during  the blue supergiant phase \citep{Lamers2001} are observed around many LBVs and cLBVs (see for example \citet{MartayanLobel2016}).  HD168625, which has the spectrum similar to MT304 \citep{Klochkova2004No12}  and, like MT304, is a BHG/cLBV star, also has a nebula around with size  $0.13\times0.17$~pc \citep{Weis2011}.  Strong circumstellar extinction is not seen for most of the Galactic LBV stars, or it is difficult to separate from interstellar extinction, as in the Pistol star case, which is completely obscured in optical range. On the other hand, there are cases such as the Homunculus nebula around $\eta$~Car, which absorbs 2~mag and has size of 0.2~pc \citep{Davidson1997,Weis2011}.    Therefore, the extinction seen in MT304 may also be partially (up to 2 mags) of circumstellar nature. Probably, the nebula around MT304 might be seen on direct images, but strong interstellar extinction and the proximity of other stars hinder this. The argument against the circumstellar origin of the reddening, however, is the absence of any evident infrared excess in the multiwavelength spectrum of MT304 \citep{Clark}, which is typical for circumstellar shells.  Moreover, the survival of such a nebula in a dense stellar cluster around MT304 is also an open question\footnote{ We may, however, note here that all the stars except for MT304 itself are low-luminosity B-dwarfs which may not easily evaporate the dust on a scales of tenths of parsec.}.
        Therefore, the exact location of the material responsible for the extra reddening is still not clear.
         
\section{Conclusions}\label{sec:concl}

    To study the high reddening affecting the hypergiant Cyg~OB2~\#12 (MT304) we conducted longslit spectroscopy and photometry of 24 stars with V=13-20~mag laying within 2.5 arcmin from this hypergiant. For 22 of these stars the spectroscopy was performed for the first time. 
    We confirmed spectroscopically {\color{black} the membership of}  MT282 and MT333, previously {\color{black} classified as members} photometrically by  \citet{MT91},  as well as for AFM27, 79 and 91 {\color{black} which have been} previously selected by X-ray emission \citep{AFM}.
    Our spectroscopic analysis also demonstrated that five more stars, not included previously, are {\color{black} also} members of the association. Thus, the extended the list of Cyg~OB2 massive stars by ten more B stars.
    
    Spectral analysis shows  that:

\begin{itemize}
\item  only 13 of the studied stars, along with MT304,   belong to Cyg~OB2. 

\item the stars  MT294, MT297, MT309  and MT340 selected photometrically by  \citet{MT91} are not members of the association, being in the foreground instead.

\item interstellar extinction increases while approaching to MT304.

\item  J203240.35+411420.1 and J203239.90+411436.2 are the most reddened massive stars after  MT304. They are located  within 13 and 15~arcsec away from MT304 and their $A_V$ are $9.1\pm0.1$~mag and $9.02\pm0.02$~mag respectively.
\end{itemize}

     Our speckle-interferometer observations confirmed  that MT304 has a second companion, which was discovered by \citet{Caballero}. We detected its orbital motion which suggests that the orbital period of the system ranges from 100 to 200 years. We also observed for the first time a third component which is weaker than main component by 4.8 mag. 

     Our study of the spatial variability of the interstellar reddening in the vicinity of MT304 revealed its strong increase towards the star, from the mean field value of 8~mag to 9~mag for closest stars to 10~mag for MT304 itself. This excess of nearly 2~mag may be due to a  circumstellar shell with radius of some tenths of parsec, large enough to partially obscure the closest stars also.

  \begin{table*}
\caption{List of the most reddened stars in Cyg~OB2. First column is the name in Sloan Digital Sky Survey (SDSS) or United States Naval Observatory (USNO-B1) catalog, 2 -- name in catalog by \citet{MT91}, \citet{AFM} or \citet{Comeron02}. Last column gives coordinates of stars not included in SDSS catalog.  } 
\label{tab:extinction}
\begin{tabular}{llllrrr}
\hline
  \multicolumn{1}{c}{SDSS}  &\multicolumn{1}{c}{MT/AFM} &\multicolumn{1}{c}{Sp. Type} &\multicolumn{1}{c}{$A_V$, [mag]} &\multicolumn{1}{c}{B, [mag]} &\multicolumn{1}{c}{V, [mag]}&\multicolumn{1}{c}{Coordinates}\\
  \multicolumn{1}{c}{(1)}   &\multicolumn{1}{c}{(2)}    & \multicolumn{1}{c}{(3)}     & \multicolumn{1}{c}{(4)}         &\multicolumn{1}{c}{(5)}      & \multicolumn{1}{c}{(6)}    & \multicolumn{1}{c}{(7)}      \\
  \hline
                              &  \multicolumn{5}{c}{ Most reddened stars according to \citet{Wright2015}  } &     \\
   {J203311.06+411032.0}      & MT435                   & B0V                         & 7.13                     & 16.97                   & 14.78&       \\ 
   {J203051.10+412021.7}      & MT20                    & B0V                         & 7.2                      & 16.66                   & 14.48&       \\ 
   {USNO-B1.0 1308-0398134}   &      A31                & B0.5V                       & 7.2                      & 15.78                   &      & $\alpha=20:32:39.50$, $\delta=+40:52:47.0$      \\ 
   {USNO-B1.0 1312-0390113}   & MT448                   & O6V                         & 7.22                     & 15.76                   & 13.61& $\alpha=20:33:13.25$, $\delta=+41:13:28.6$     \\ 
   {J203444.10+405158.6}      &      A24                & O6.5III                     & 7.29                     & 15.3                    &      &       \\ 
   {J203039.69+410848.7}      &      A23                & B0.7Ib                      & 7.38                     &                         &      &       \\ 
   {J203323.46+410912.9}      & MT516                   & O5.5V                       & 7.43                     & 14.04                   & 11.84&       \\ 
   {J203444.69+405146.7}      &      A27                & B0Ia                        & 7.58                     &                         &      &       \\ 
   {J203547.08+412244.6}      &      WR146              & WC6+O8III                   & 7.83                     &                         &      &       \\ 
\textbf{J203231.49+411408.4}  & \textbf{MT267$^{*}$}          & \textbf{O7.5III}            & \textbf{8.03}            & \textbf{15.38}          &      &       \\
   {J203027.29+411325.2}      &      B17                & O7I+O9I                     & 8.08                     & 15.61                   &      &       \\ 
   {J203302.88+404725.1}      &      A20                & O8II                        & 8.13                     & 15.12                   &      &       \\ 
   {J203136.90+405909.0}      &      A15                & O7I                         & 8.26                     & 15.83                   &      &       \\ 
   {USNO-B1.0 1312-0390173}   & MT488                   & B2Ve                        & 8.27                     & 17.24                   & 14.88& $\alpha=20:33:18.55$, $\delta=+41:15:35.4$      \\ 
\textbf{J203240.89+411429.6}  & \textbf{MT304$^{**}$}          & \textbf{B3.5Ia+}            & \textbf{10.2}            & \textbf{14.81}          & \textbf{11.46}&       \\
                              &                         &                             &                          &                         &      &       \\   
\cline{1-2}
                              &  \multicolumn{5}{c}{ Most reddened stars according to this work  } &     \\
  {J203243.61+411513.9}      & AFM79                   &  B7 V                       & 7.0                      & 19.8                   &  17.56 &        \\ 
\textbf{J203231.49+411408.4}  & \textbf{AFM17$^{*}$}          &  \textbf{O7.5 IIf}          & \textbf{7.68}            & \textbf{15.14}         &  \textbf{12.9} &        \\
   {J203235.63+411509.6}      & AFM27                   &  B2 V                       & 7.7                      & 18.14                  &  15.82 &        \\ 
   {J203248.62+411429.8}      & MT333                   &  B1 V                       & 7.7                     & 17.85                  &  15.54 &        \\ 
   {J203235.33+411445.3}      & MT282                   &  B1 IV                      & 7.94                     & 17.29                  &  14.9  &        \\ 
   {J203238.09+411441.8}      &                         &  B5 V                       & 8.4                     & 20.85                  &  18.24 &        \\ 
   {J203241.80+411439.2}      &                         &  B7 V                       & 8.66                      & 21.7                   &  18.94 &        \\ 
   {J203239.90+411436.2}      &                         &  B5 V                       & 9.03                      & 21.58                  &  18.74 &        \\ 
   {J203240.35+411420.1}      &                         &  B2 V                       & 9.15                      & 20.57                  &  17.76 &        \\ 
\textbf{J203240.89+411429.6}  & \textbf{MT304$^{**}$}          &  \textbf{B5 Ia-0}           & \textbf{10.0}           & \textbf{14.76}         & \textbf{11.5}&        \\
\\
\cline{1-2}
\multicolumn{7}{l}{$^{*}$ -- MT267 is included in our work with name AFM17. }\\
\multicolumn{7}{l}{$^{**}$ -- for MT304 and  MT267 our spectral classification differs from  \citet{Wright2015}.}\\
\hline
\end{tabular}
\end{table*}
 
\section*{Acknowledgments}
We would like to thank  the anonymous referee for a helpful and insightful report. The observations at the Russian 6-meter telescope were carried out with the financial support of the Ministry of Education and Science of the Russian Federation (agreement No. 14.619.21.0004, project ID RFMEFI61914X0004). The study was supported by the Russian Foundation for Basic Research (projects no. 14-02-31247,14-02-00291,13-02-00419, 14-02-00759). Olga Maryeva thanks the grant of Dynasty Foundation. The work is performed according to the Russian Government Program of  Competitive Growth of Kazan Federal University. Sergey Karpov thanks the grant of Russian Science Foundation No~14-50-00043.

\bibliographystyle{mn2e}
\bibliography{Maryevabib}

\section*{Appendix A}
Spectral data.
\begin{figure*}
\includegraphics[scale=0.35]{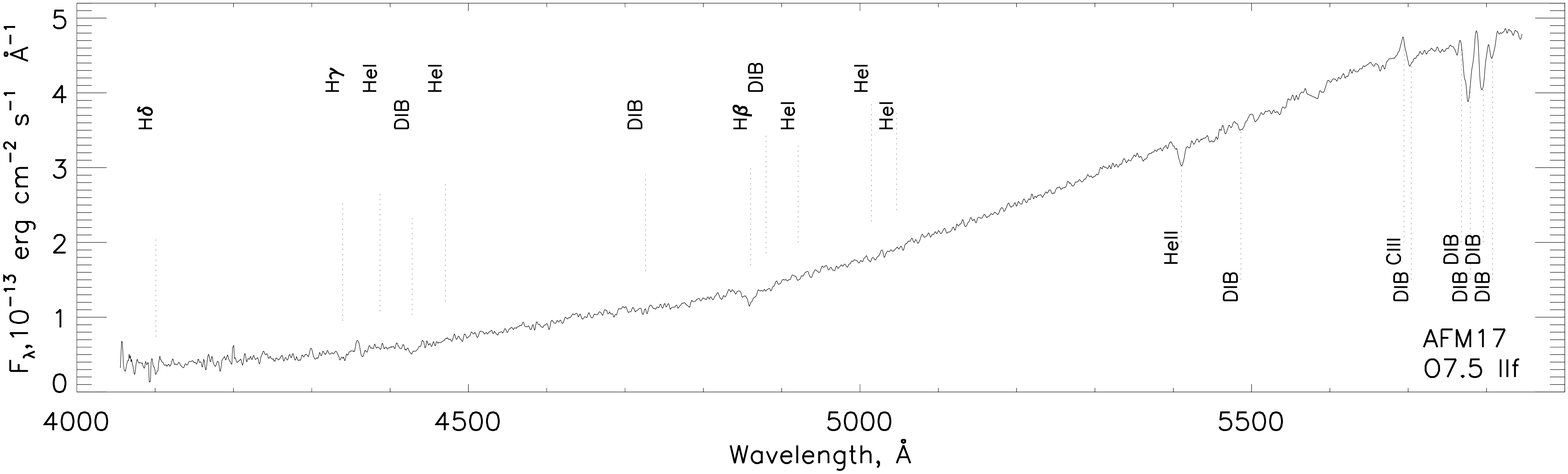} 
\includegraphics[scale=0.35]{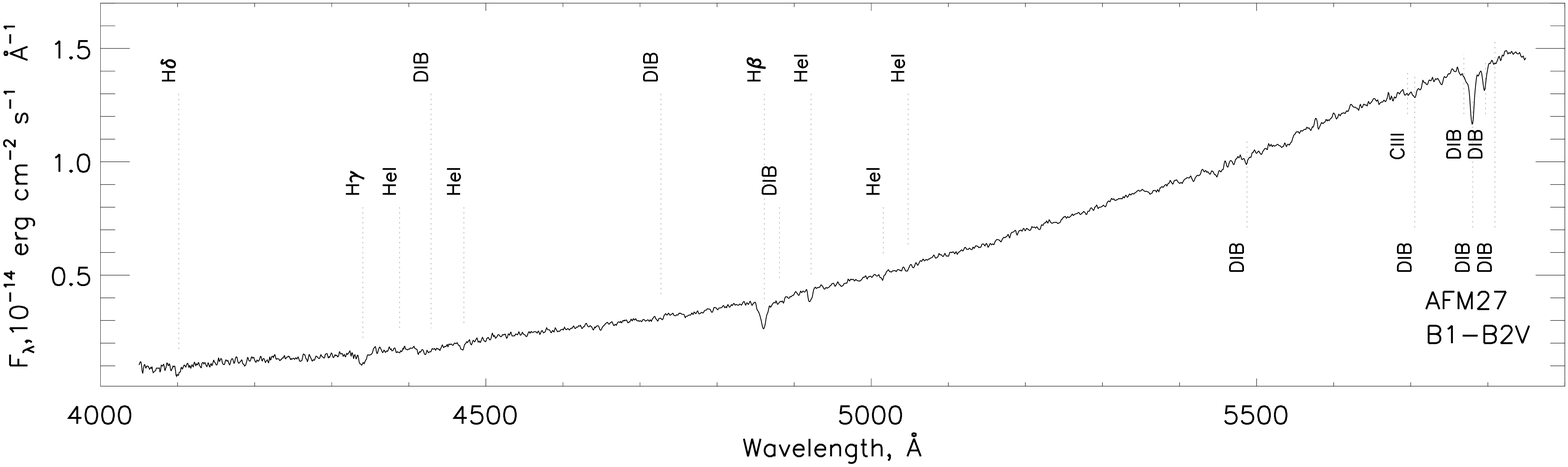}
\includegraphics[scale=0.35]{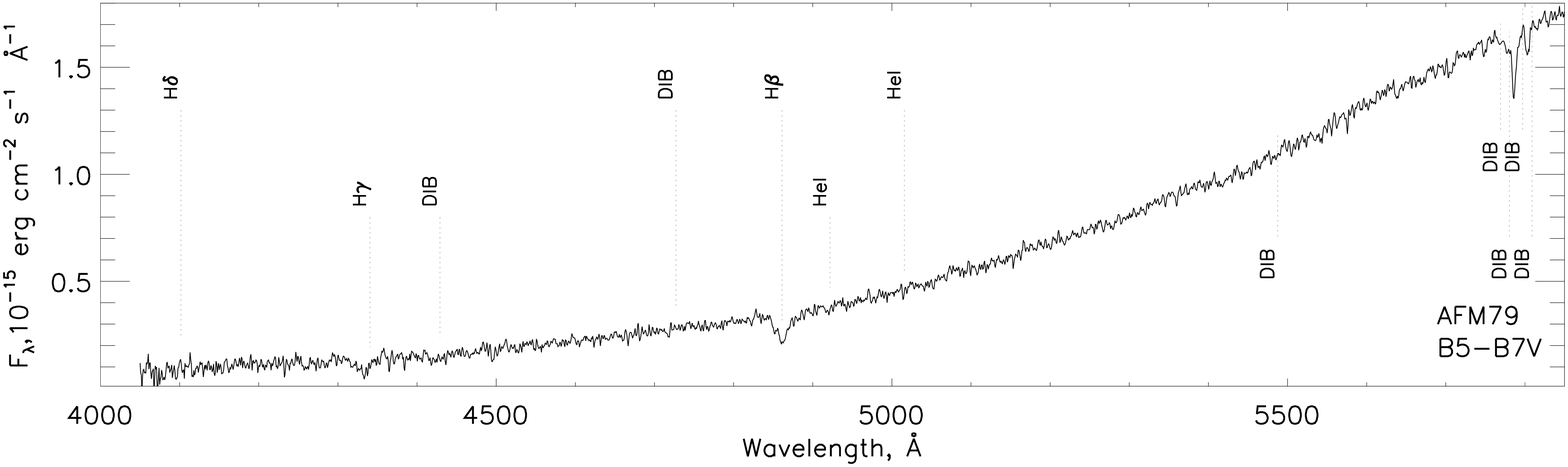}
\includegraphics[scale=0.35]{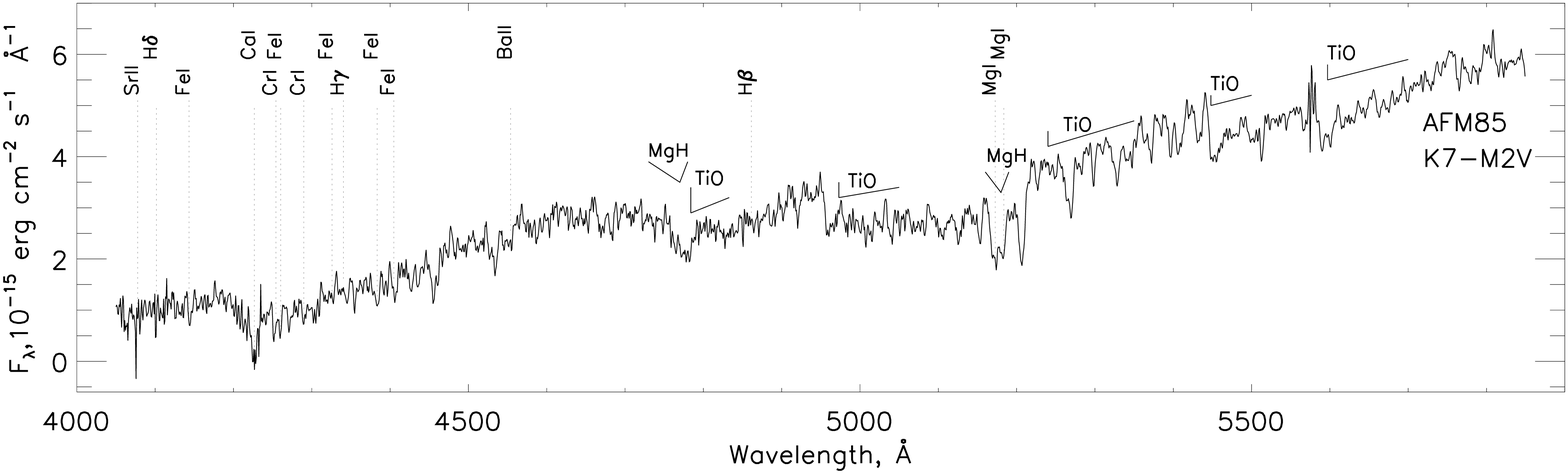}
\caption{\color{black}  From the top panel downward, spectra of the stars: J203231.49+411408.4, J203235.63+411509.6, J203243.61+411513.9 and J203243.49+411445.1. }
\label{fig:spectrum17}
\end{figure*}
\begin{figure*}
\includegraphics[scale=0.35]{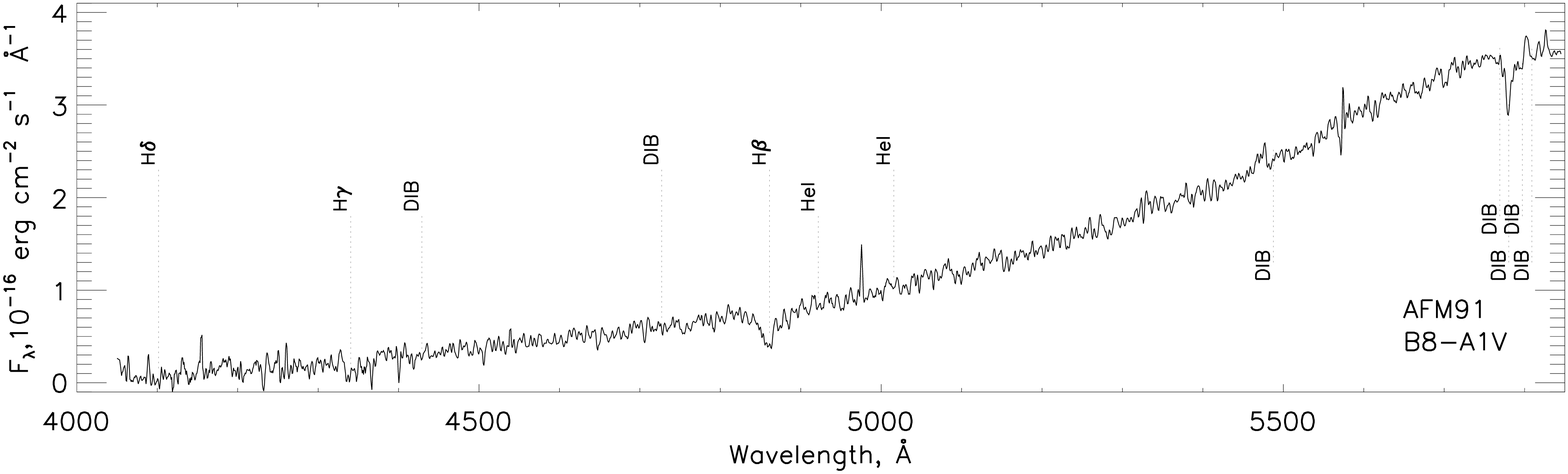}
\includegraphics[scale=0.35]{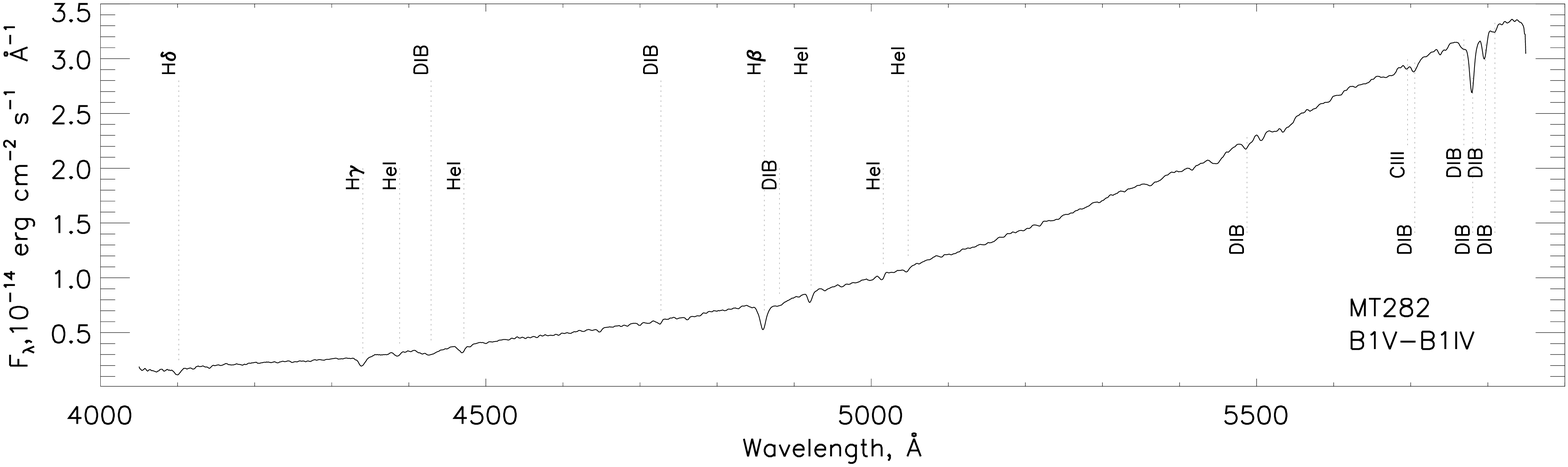}
\includegraphics[scale=0.35]{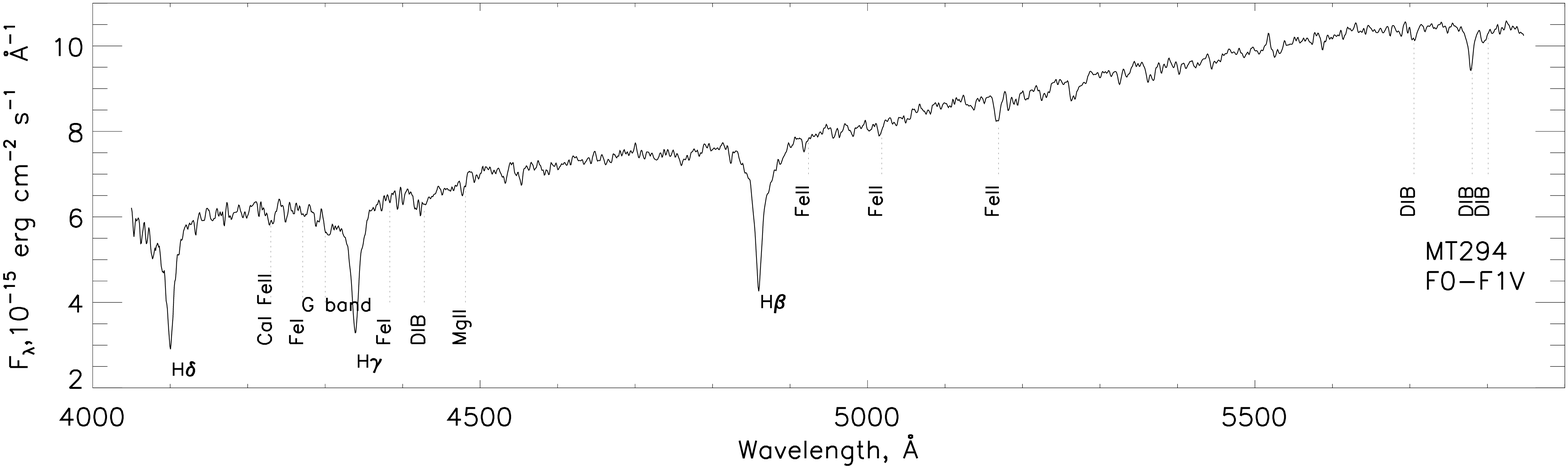}
\includegraphics[scale=0.35]{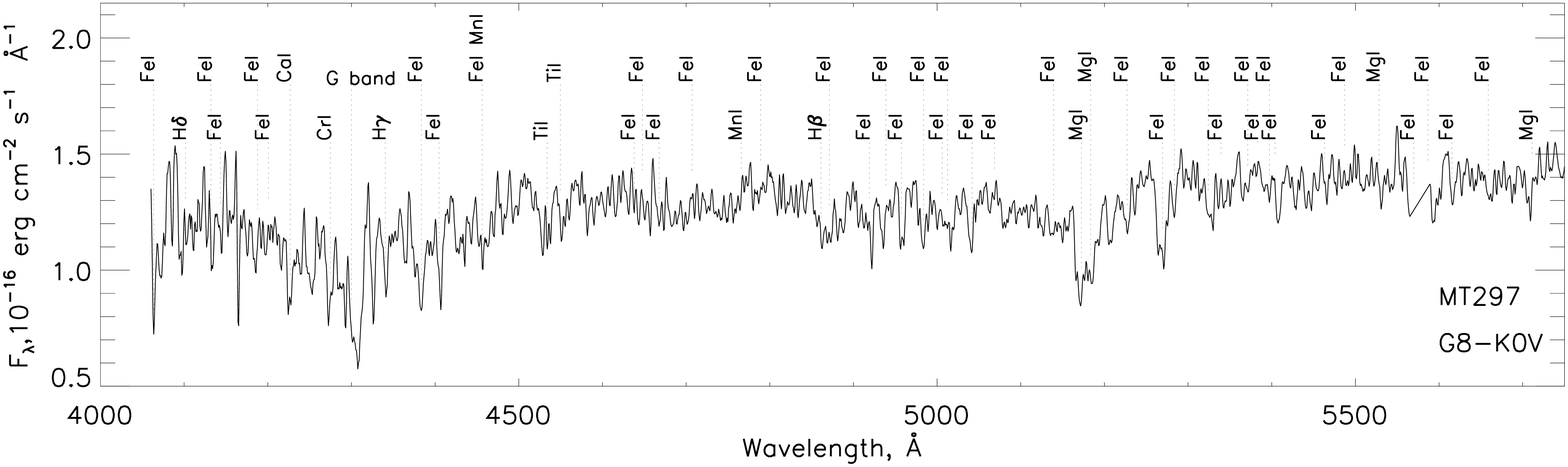}
\caption{\color{black}  From the top panel downward, spectra of the stars: J203245.07+411416.5, J203235.33+411445.3, J203237.89+411509.7 and J203235.63+411509.6.}
\label{fig:spectrum297}
\end{figure*}

\begin{figure*}
\includegraphics[scale=0.35]{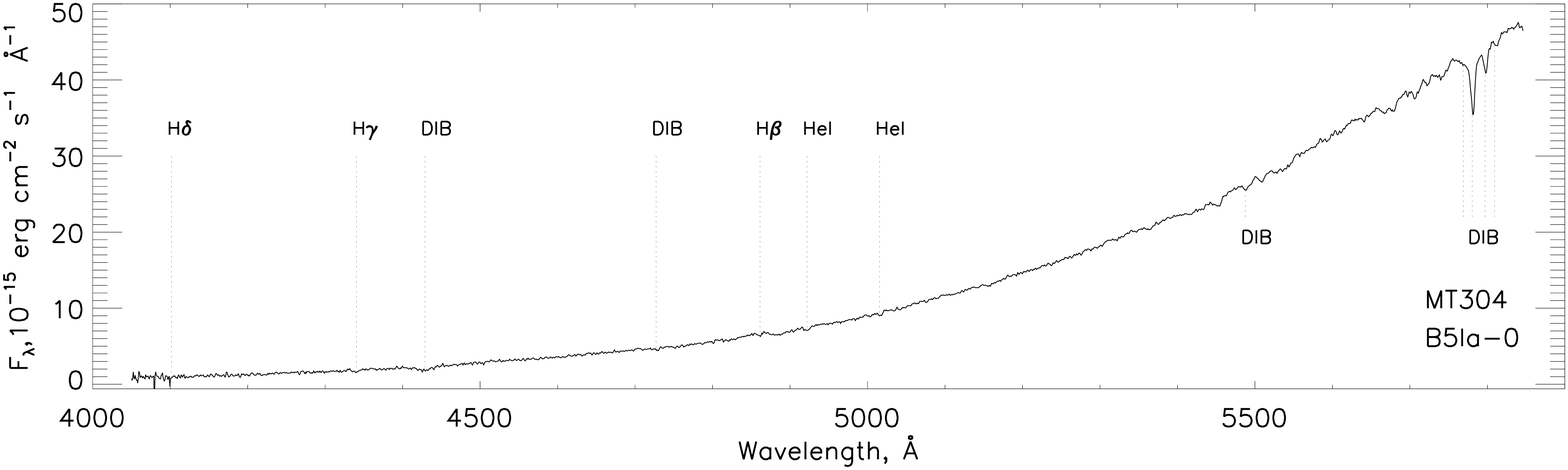}
\includegraphics[scale=0.35]{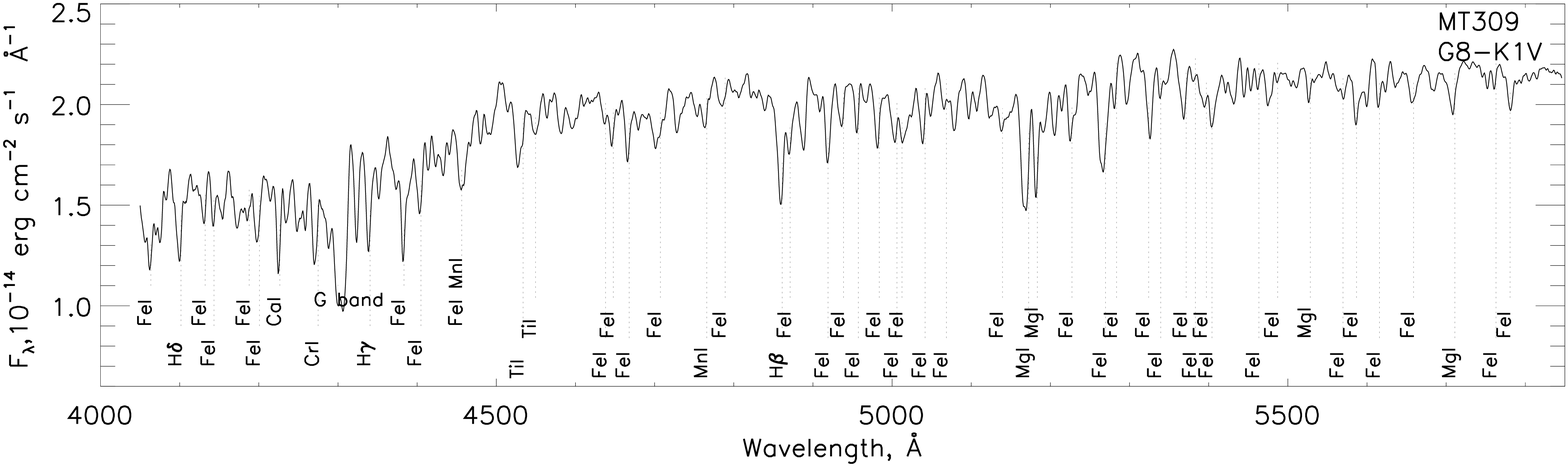}
\includegraphics[scale=0.35]{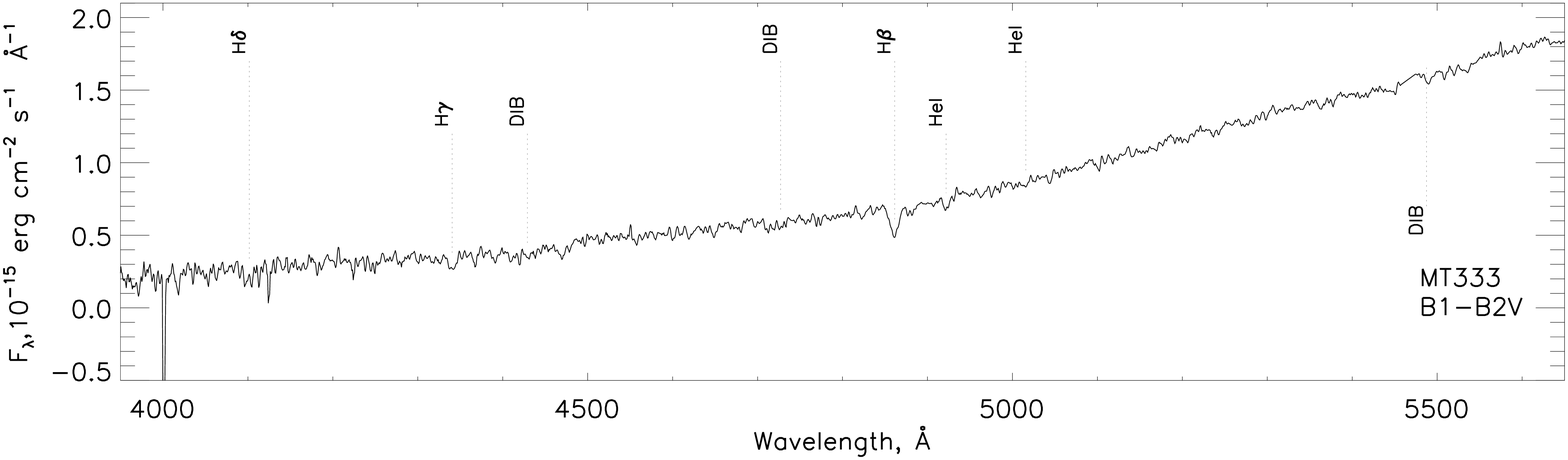}
\includegraphics[scale=0.35]{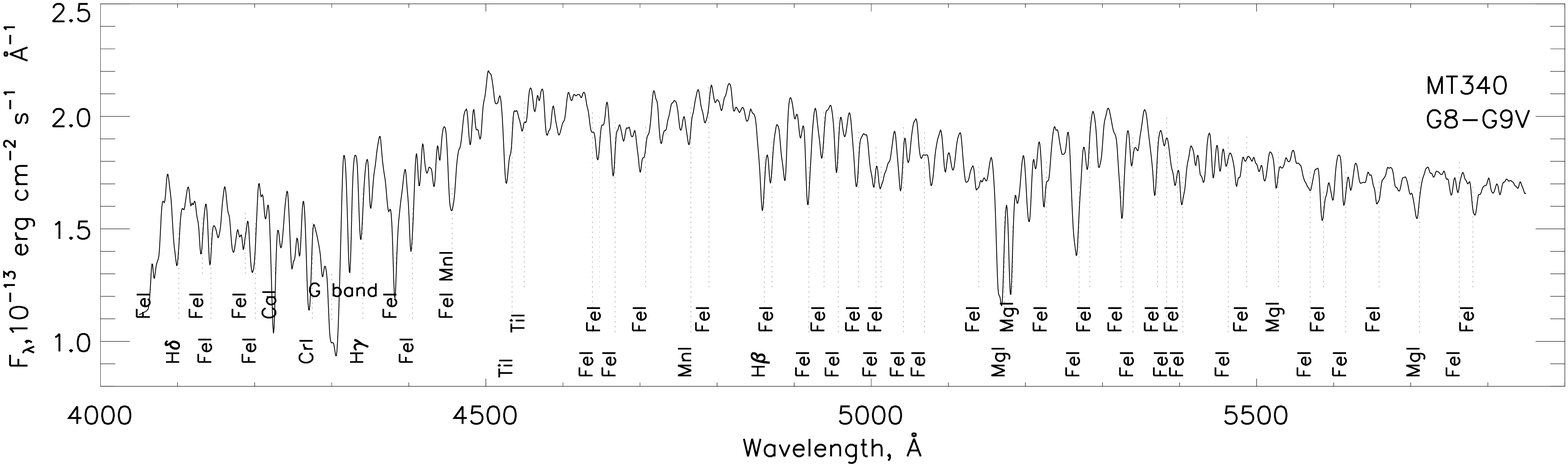}
\caption{\color{black} From the top panel downward, spectra of the stars: J203240.89+411429.6, J203241.84+411422.14, J203248.62+411429.8 and J203250.25+411448.4.}
\label{fig:spectrum309}
\end{figure*}

\begin{figure*}
\includegraphics[scale=0.35]{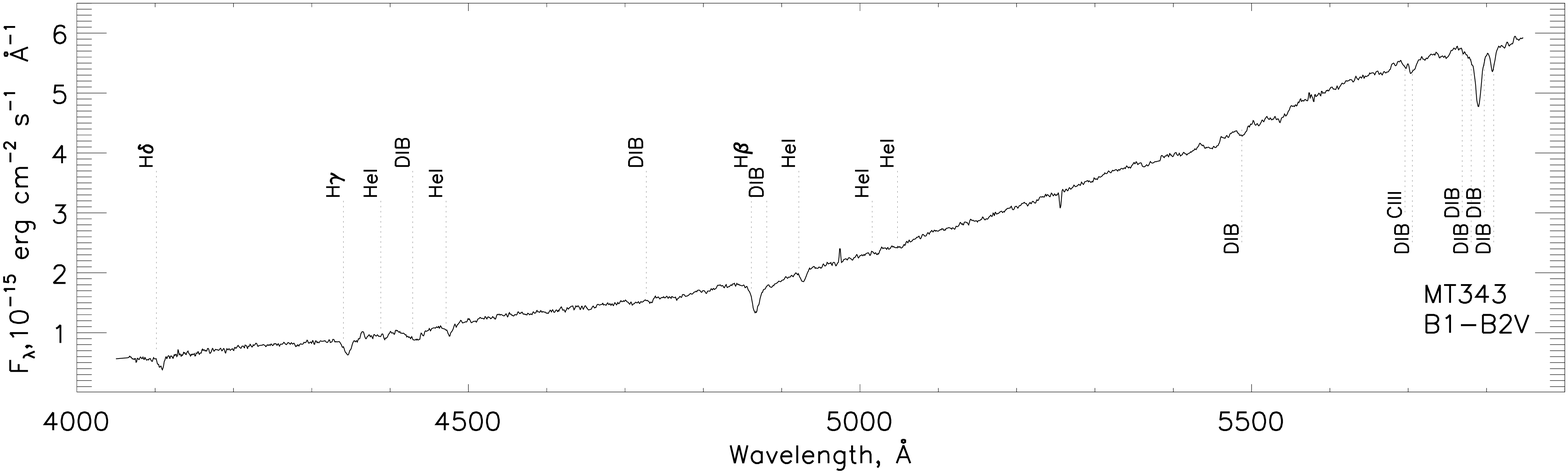}
\includegraphics[scale=0.35]{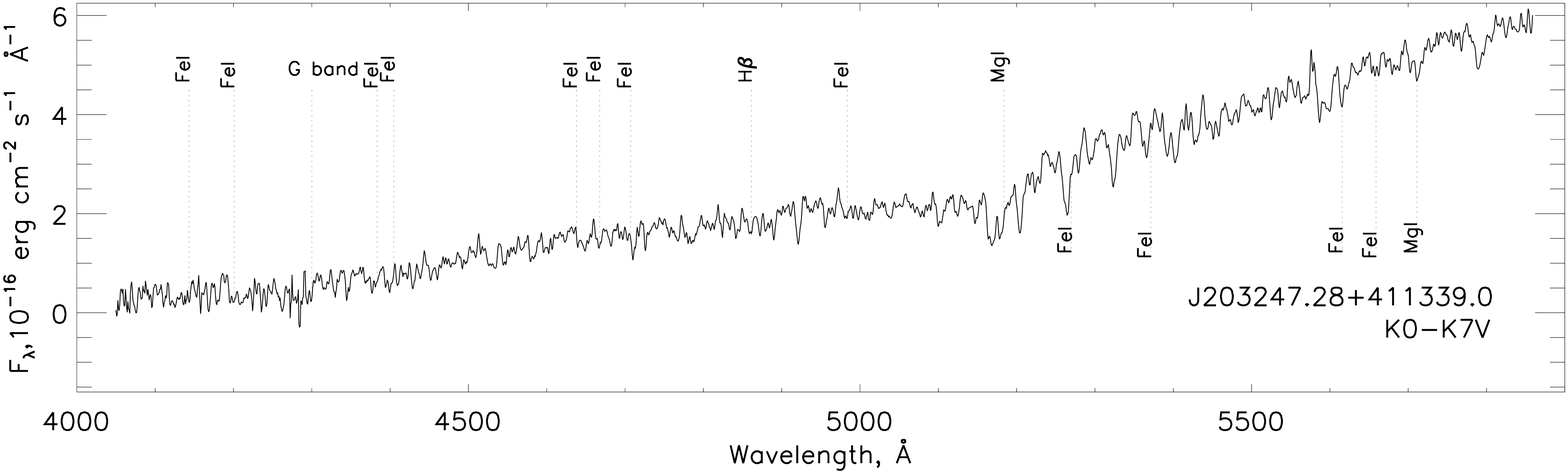}\\
\includegraphics[scale=0.35]{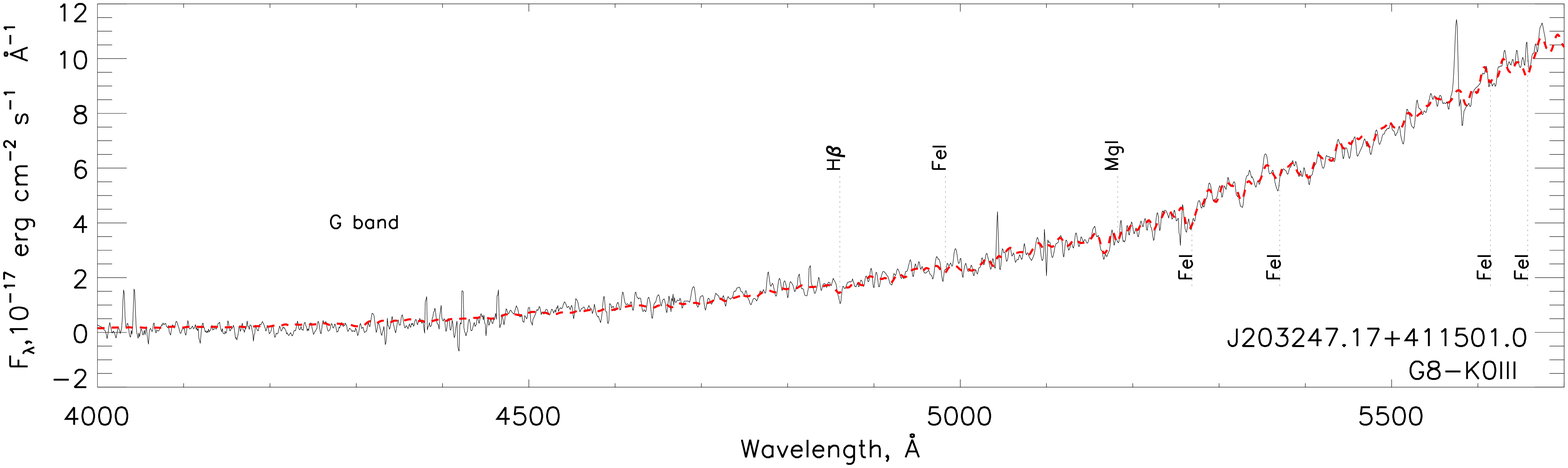}\\ 
\includegraphics[scale=0.35]{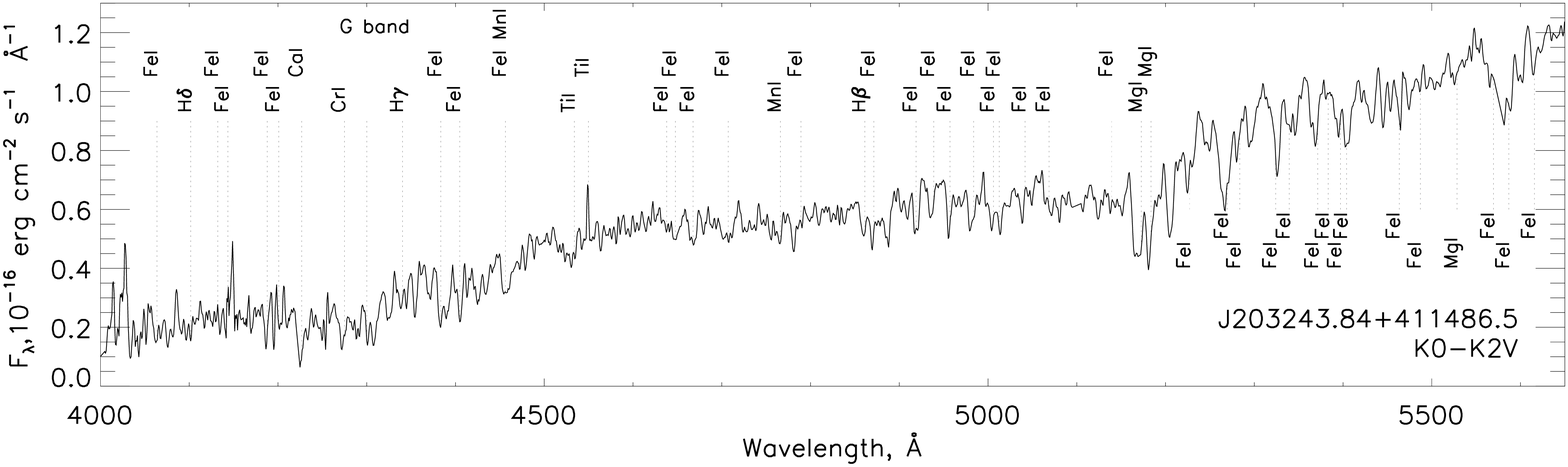}
\caption{\color{black}  From the top panel downward, spectra of the stars: J203250.75+411502.2, J203247.28+411339.0, J203247.17+411501.0 and J203243.84+411446.5. 
         For comparison the reddened spectrum of HD113226 (G8~III) is shown in the third panel by dashed line.}
\label{fig:spectrum343}
\end{figure*}
\begin{figure*}
\includegraphics[scale=0.35]{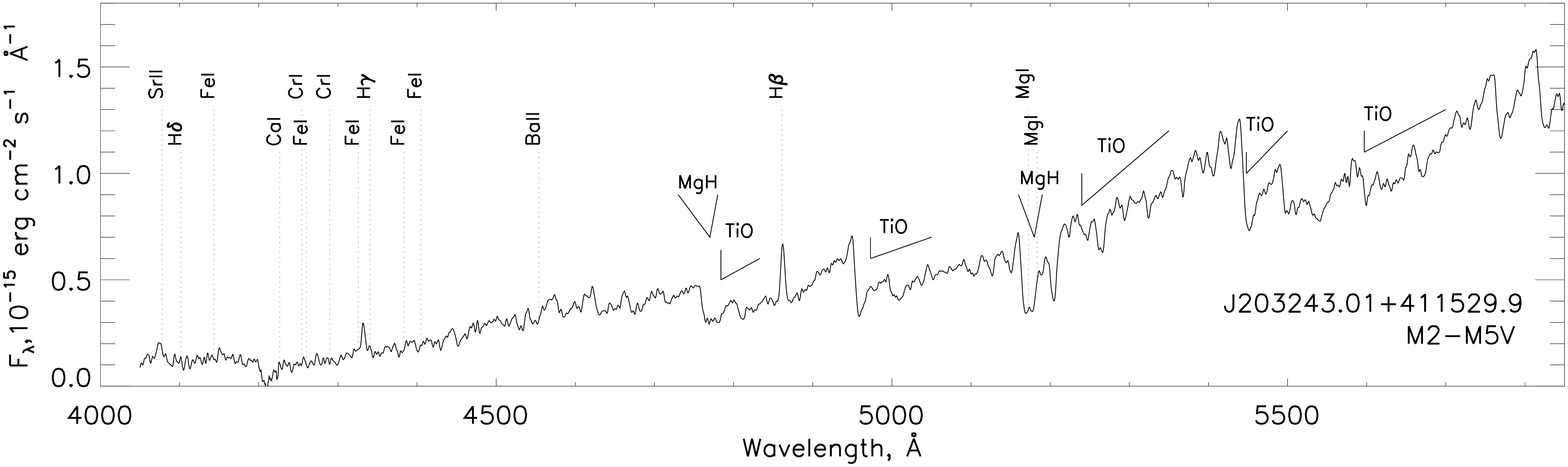}\\ 
\includegraphics[scale=0.35]{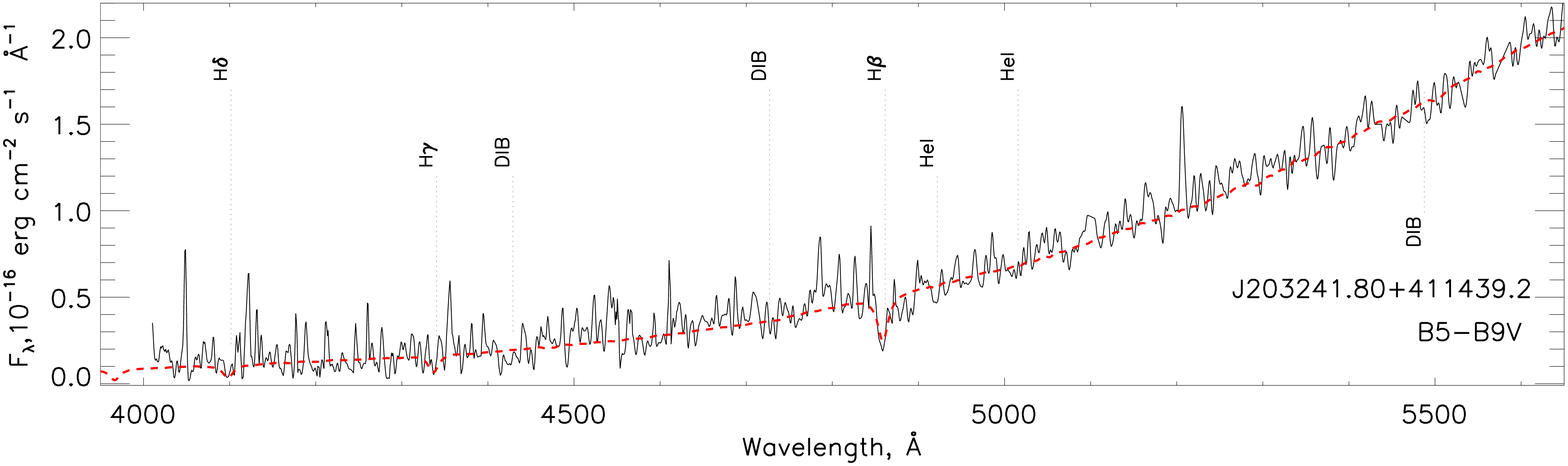}
\includegraphics[scale=0.35]{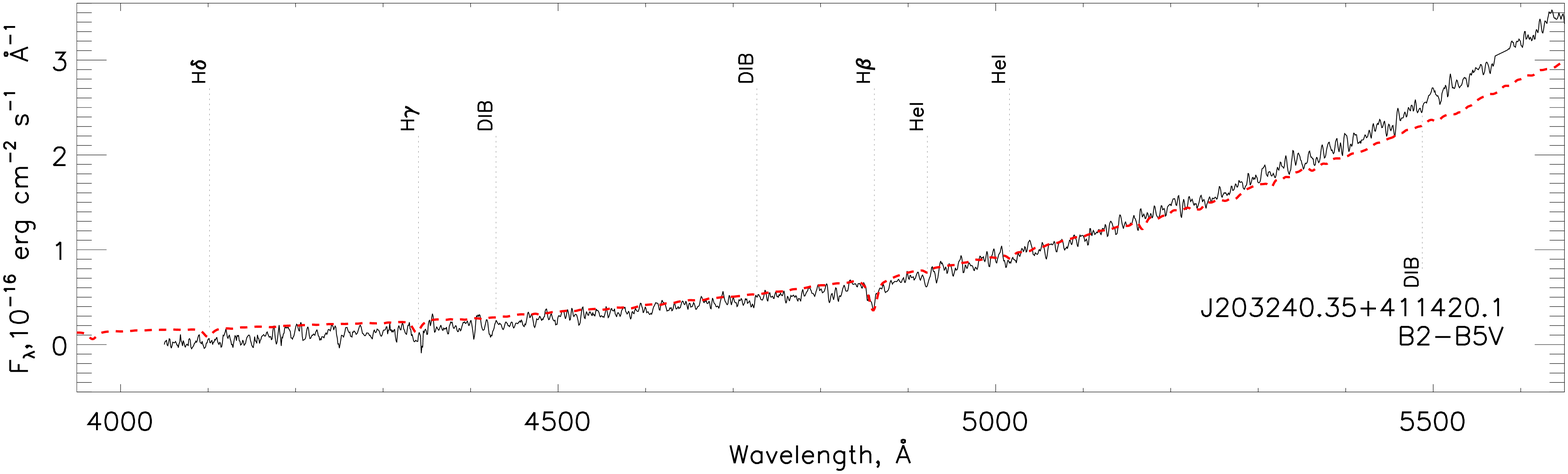}
\includegraphics[scale=0.35]{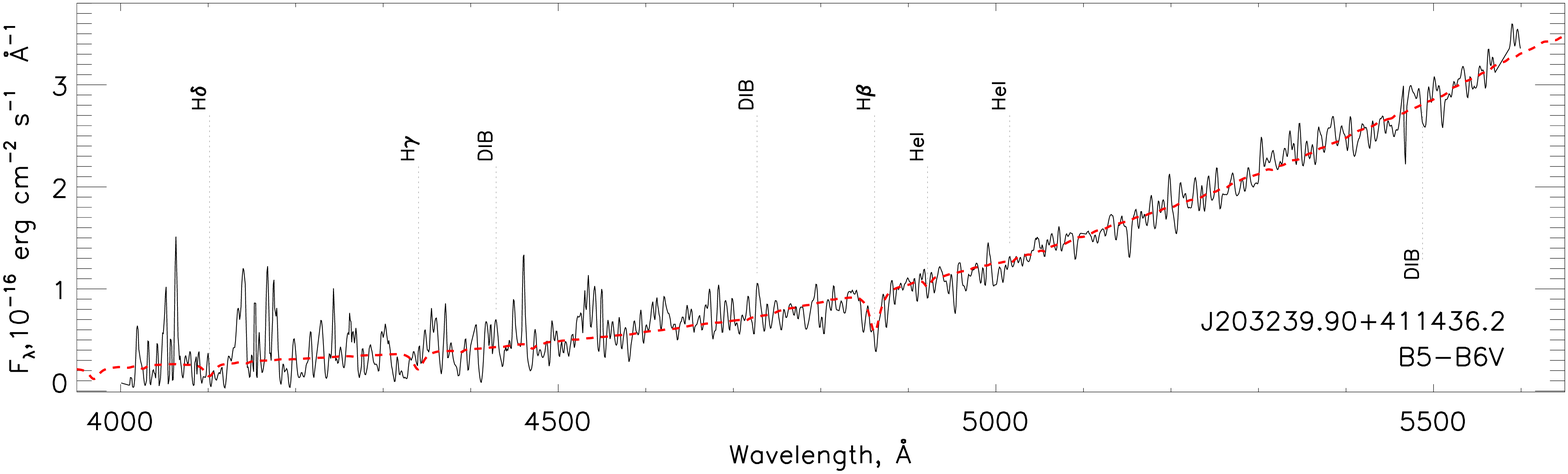}
\caption{\color{black}  From the top panel downward, spectra of the stars: J203243.01+411529.9, J203241.80+411439.2, J203240.35+411420.1 and J203239.90+411436.2. 
         For comparison the reddened spectra of HD144206 (B9~III), HD39866 (A2~II) and HD147394 (B5~IV) are shown  by dashed lines.}
\label{fig:spectrumj90}
\end{figure*}
\begin{figure*}
\includegraphics[scale=0.35]{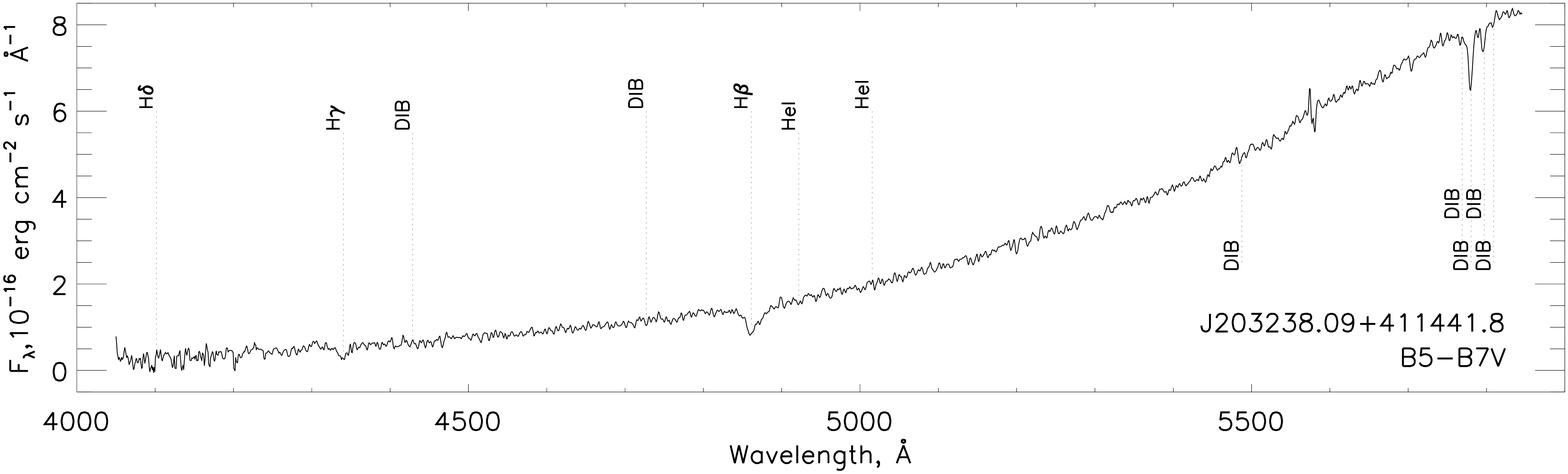}
\includegraphics[scale=0.35]{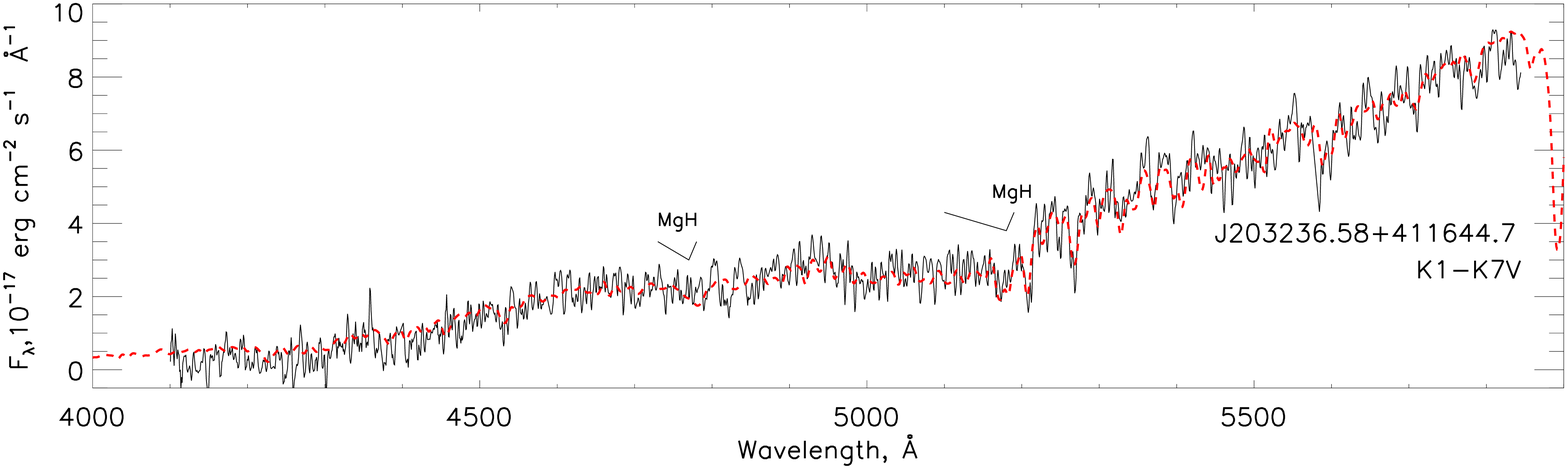}\\
\includegraphics[scale=0.35]{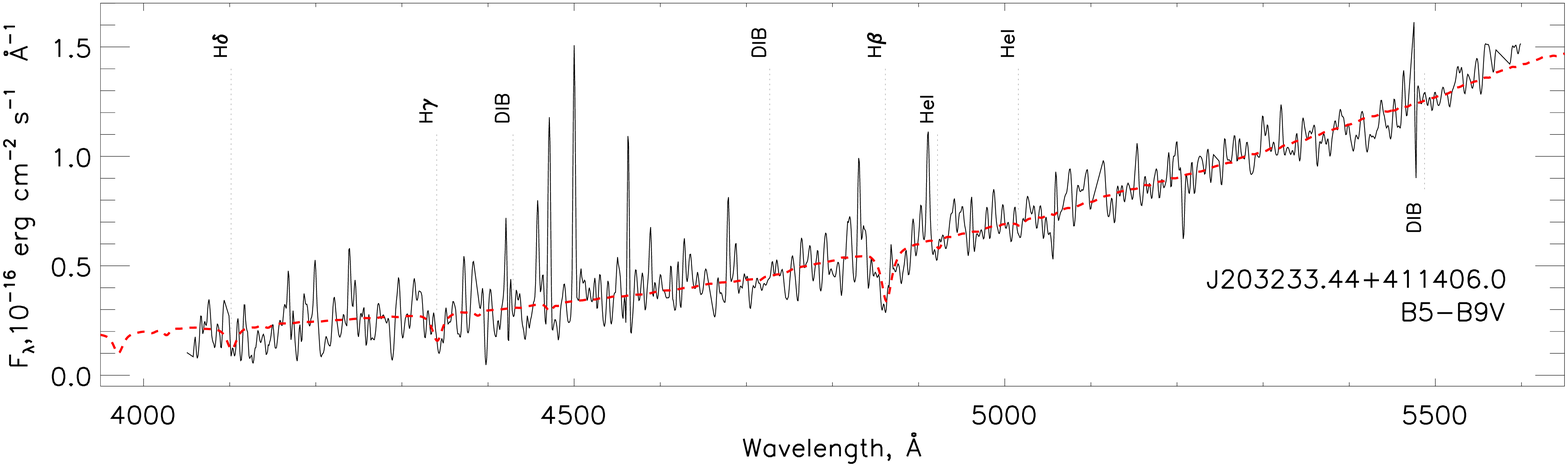}
\includegraphics[scale=0.35]{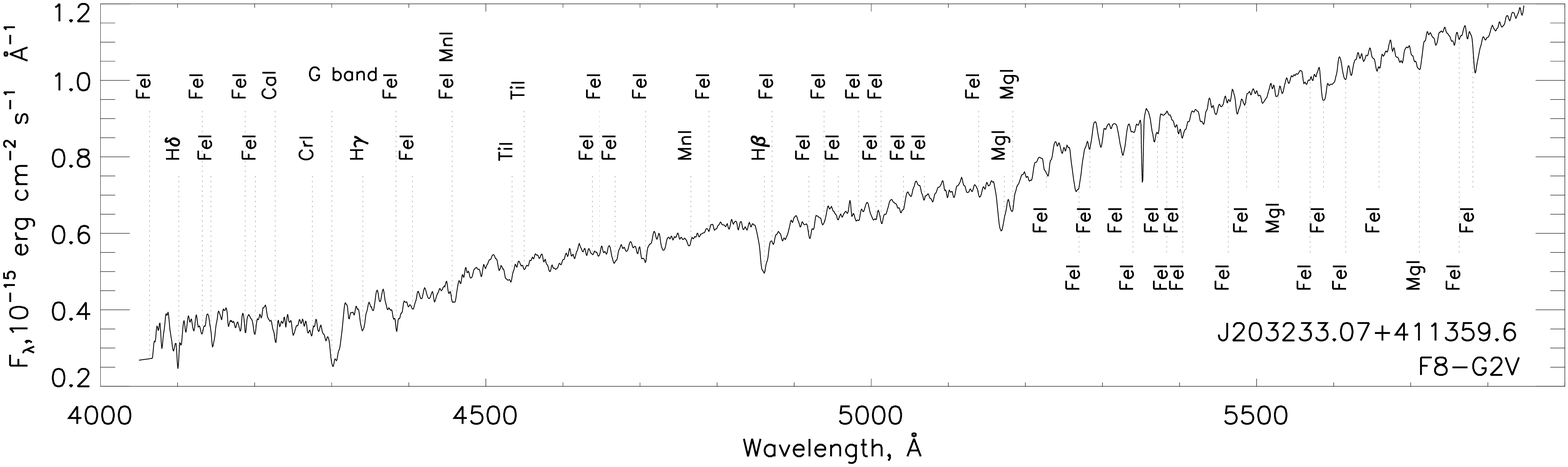}
\caption{\color{black}   From the top panel downward, spectra of the stars: J203238.09+411441.8, J203236.58+411644.7, J203233.44+411406.0 and J203233.07+411359.6.
         For comparison the reddened spectra of HD157881 (K7~V) and HD147394 (B5~IV) are shown  by dashed lines. }
\label{fig:spectrumj07}
\end{figure*}

\begin{figure*}
\includegraphics[scale=0.35]{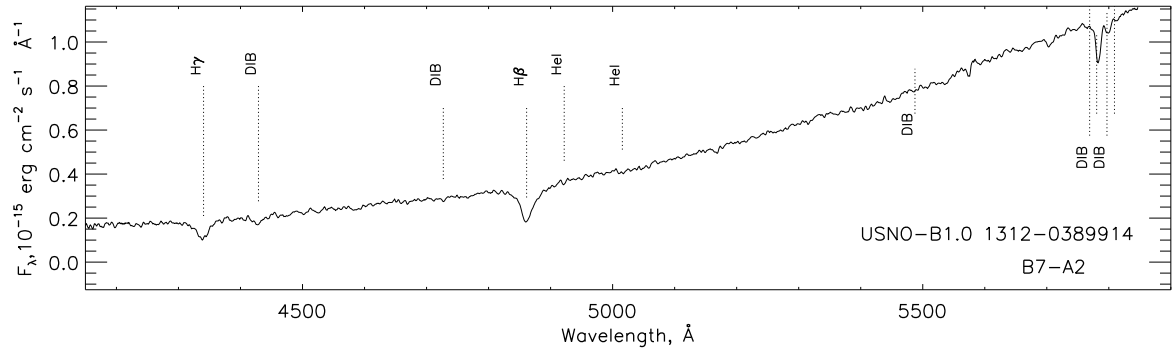}
\caption{Spectrum of  USNO-B1.0 1312-0389914.}
\label{fig:spectrumusno}
\end{figure*}

\label{lastpage}

\end{document}